%% file: main.tex


\documentclass[fullpaper,final]{nldl}

\usepackage[T1]{fontenc}
\usepackage[utf8]{inputenc}
\usepackage[backend=biber,style=numeric-comp]{biblatex}
\usepackage[english]{babel}
\usepackage{textcomp}
\usepackage{graphicx}
\usepackage{enumitem}
\usepackage{csquotes}
\tcbuselibrary{breakable}
\usepackage{caption}
\usepackage{makecell} 
\usepackage[hidelinks, hypertexnames=false]{hyperref}
\usepackage{booktabs}
\usepackage{makecell}
\usepackage{listings}

\usepackage{hyperref}
\usepackage{url}
\hypersetup{
  colorlinks,
  linkcolor = BrickRed,
  citecolor = NavyBlue,
  urlcolor  = Magenta!80!black,
}

\title{title}
\author[]{Liv Hilde Sjøflot}
\author[]{Tobias A. Opsahl}
\affil[]{University of Oslo}
\affil[]{\texttt{livhildes@gmail.com, tobiasao@uio.no}}

\begin{document}
\title{Deceptive Cookies: Consent by Design - A Mixed Methods Study}

\maketitle
    
\begin{abstract}
    While companies increasingly rely on data, especially when it comes to targeted advertising, adapting content to users, selling data and training machine learning models, the collection of data raises privacy concerns. One way of collecting data is by using HTTP cookies when interacting with a website. Legal regulations require service providers to collect consent for some forms of cookie collection, which is often acquired through \emph{cookie consent banners}, but their effectiveness has been debated. This study aims to understand what influences users' experience and behaviour when managing their cookie consent, by investigating the gap between their stated privacy preferences and their actual actions. A mixed methods approach was used, collecting data from a usability test and a survey (N=20). The results showed that although participants generally want to reject cookie collection, they often end up accepting because of deceptive patterns in the cookie consent banner design. It also showed that they were more willing to consent to websites they trusted and if they expected it would improve their user experience. Although the current EU legislation states that withdrawing consent must be as easy as giving it, withdrawing consent took on average more than 20 times longer than giving it. This suggests that cookie consent banners in their current form are not ideal with respect to user autonomy, often leading users to \emph{consent by design}.
\end{abstract}

\input{01introduction}
\input{02background}
\input{03methodology}
\input{04analysis_and_results}
\input{05discussion}
\input{06conclusions_and_future_work}
\input{07limitations}

\printbibliography

\input{08appendix}

\end{document}

%% file: 01introduction.tex
\section{Introduction}
\label{sec:introduction}

Data has substantial value in the current digital society, where businesses are increasingly data-driven \cite{2023age_of_surveillance_capitalism}. For instance, gathering user data can provide business value by understanding users' behaviour, needs and backgrounds, which can be used to improve and personalise services. User data is also heavily used in the marketing industry, which allows for the personalisation of both the content and the placements of advertisements, targeting users who are believed to be more inclined to purchase certain products or services based on their data \cite{2025ad_spending_forecast}. Artificial intelligence and machine learning methods have recently grown in popularity, which often require enormous amounts of training data \cite{2022laion-5b, 2023gemini, 2024llama}. Companies are increasingly using these technologies in order to increase business value, and as a result, vast amounts of data are being collected in order to train and utilise these models.

As a large amount of the data collected is personal and about individual users, such as text, images, videos and online behaviour patterns, this data collection raises privacy concerns. The concept of privacy has many definitions, but it generally refers to the individual's ability and right to exercise control over their own data \cite{1890right_to_privacy, 1949universal_human_rights, 2000european2_fundamental_rights}. In spite of multiple laws and legal requirements, such as the \emph{General Data Protection Regulation} (GDPR) regulating data protection, data subjects' rights and consent \cite{GDPR}, there is an abundance of cases where users' personal data and privacy rights are being violated.

For instance, the dating application \emph{Grindr} was fined by the Norwegian Data Protection Agency (Datatilsynet) for selling and sharing their users' personal data, including their HIV-status, to third party advertising companies \cite{grindr_datatilsynet}. The company \emph{Clearview AI}, which uses facial recognition models reportedly trained on over 50 billion images \cite{times_clearview_ai}, is facing multiple investigations, lawsuits and fines for GDPR violations and insufficient cooperation with the supervisory authority \cite{pipeda_clearview_ai_investigation, dutch_fine_clearview_ai, several_fines_enformcementtracker}. Google Analytics, which is used for providing insight and statistics to websites, such as user behaviour and demographics, was deemed by the Norwegian Data Protection Agency to have been illegally storing personal data from European websites outside the European Economic Area \cite{datatilsynet_google_analytics}. However, the rules for data transfers have since changed, making Google LLC and Google Analytics exempt from some of the privacy regulation \cite{datatilsynet_data_transfer, data_privacy_framework}.

One common method of data collection is through \emph{HTTP cookies}, which have become central to online tracking. To comply with data collection regulations, websites use \emph{cookie consent banners}. However, these banners often fail to meet the conditions for valid consent and employ \emph{deceptive patterns} (or \emph{dark patterns}) \cite{2010dark_patterns} that hinder data subject autonomy and push users toward consent \cite{take_some_cookies_degeling2018, GDPR&cookieBannerImpact_kretschmer2021}. This creates a gap between individuals’ stated privacy preferences and their actual behaviour.

This work investigates that gap by examining how users interact with cookie consent banners, what influences their consent decisions and whether regulatory requirements for withdrawing consent are met. A mixed methods design was used, combining usability testing with a think-aloud protocol and a survey of 20 participants in Norway. The study addresses the following research questions (RQ):

\begin{enumerate}[label=\textbf{RQ}\textbf{\arabic*}]
    \itemsep0em 
    \item How do users respond to various cookie consent banners?
    \item What are users’ general privacy preferences and how do they compare to their responses to the cookie consent banners?
    \item Is withdrawing consent as easy as giving it?
    \item What are users’ general feelings, attitudes and preferences with regards to cookie consent banners?
\end{enumerate}

This study makes the following contributions:

\begin{itemize}
    \itemsep0em 
    \item New empirical data from usability tests.
    \item Quantitative evidence showing that banner design strongly affects users' decisions and that withdrawing consent takes on average more than twenty times longer than giving it.
    \item Qualitative insights into users' motivations, attitudes and feelings toward cookie consent. 
\end{itemize}

Together, these findings highlight the tension between regulatory intent and practical user experience, suggesting that current consent mechanisms often lead to consent by design rather than by choice.

%% file: 02background.tex
\section{Background}
\label{sec:background}

\subsection{HTTP Cookies}
\label{subsec:http_cookies}

Hypertext transfer protocol (HTTP) cookies are small text files of key-value pairs stored on user agents by web servers, where each pair is a single cookie \cite{barth2011rfc}. The HTTP protocol is stateless, but cookies allow web servers to communicate stateful information. \emph{Session cookies} exist only for the duration of a session, while \emph{persistent cookies} can exist beyond a session and have an explicit expiration date \cite{bujlow2017survey}. Cookies that are essential for providing the service are referred to as \emph{strictly necessary cookies} \cite{2002eprivacy}, while \emph{third-party cookies} are set by external services, such as advertisements \cite{barth2011rfc}.

\subsection{The right to privacy} 

In 1890, ``the right to privacy'' was referred to as ``the right to be left alone'' \cite{1890right_to_privacy}, and has since then been an important legal concept. It is mentioned in Article 12 of \emph{The Universal Declaration of Human Rights}, established in 1948 \cite{1949universal_human_rights}. In the European Union (EU), the \emph{Charter of Fundamental Rights} of the EU regulates privacy in Article 7 covering ``respect for private and family life'' and in Article 8 covering the ``protection of personal data'' \cite{2000european2_fundamental_rights}. In addition, the \emph{ePrivacy Directive} and its amendments regulate information privacy in electronic communications \cite{2002eprivacy}. Its 2009 amendment, which came into force in 2011, requires prior and informed consent from the users before initialising non-essential cookies \cite{2009eprivacy}. As a result, cookie consent banners became a common component of most websites targeting EU citizens \cite{take_some_cookies_degeling2018}.

Personal data is also protected in the EU by the GDPR since 2018, which among other things sets requirements for what a valid consent to data collection is \cite{GDPR}. Article 7 states that for a consent to be valid, it must be freely given, specific, informed and unambiguous. These requirements are further specified in recital 32 to the GDPR, specifying that consent obtained through passive means such as pre-selection is not to be considered valid, and that an electronic consent request must be ``clear, concise and not unnecessarily disruptive'' \cite{recital_32_GDPR}. In addition, Article 7 of the GDPR specifies that withdrawing consent must be as easy as giving it. 

The Norwegian \emph{law of electronic communications} (ekomloven) among other things regulates the use of cookies and what valid consent is \cite{ekomloven}. The Norwegian Data Protection Authority (Datatilsynet) has published a guide for achieving valid consent \cite{guide_datatilsynet}. This guide specifies that access to a service shall not be withheld until consent is given, that rejecting consent shall not require more clicks or steps than accepting consent, and that the option to reject should be visually equal to the option of accepting consent.

\subsection{Deceptive patterns}

The term \emph{deceptive patterns}, also known as \emph{dark patterns} as the term was first introduced \cite{2010dark_patterns}, can be understood as design choices that exploit users to make them behave in a way that is beneficial to a service, but go against the interests of the users \cite{brignull-2023}. In \textcite{nouwens2020dark}, which scraped the cookie consent banners of 10000 websites, it was found that only 12.6\% of the websites had a ``reject all'' option available within the same amount of clicks as the ``accept'' option. A field study published in 2019 showed that the use of \emph{nudges} \cite{nudge}, such as highlighting the accept option in the cookie consent banner, greatly increased the chance of users accepting consent requests \cite{utz2019informed}. \textcite{this_website_uses_cookies} investigated users' perceptions and reactions to cookie disclaimers, finding that users often did not perceive them as useful and that the level of trust they had in the website influenced their decisions.

Multiple studies have shown that there exists a contradiction between users' reported privacy preferences and their actual behaviour related to privacy. Users express a general concern for their privacy, but may still disclose and share their data, for instance in the form of accepting cookies. This dissonance between privacy preferences and behaviour has been referred to as the \emph{privacy paradox} \cite{norberg2007privacy}. However, as shown by \textcite{actual_behaviour_2017Kokolakis}, there is an ongoing discussion as to whether or not this privacy paradox actually exists. The article \textcite{solove2021myth} criticises the privacy paradox perspective, which blames users for being irrational by giving up their personal data for little to no value and rather focuses on holding businesses accountable by addressing deceptive designs. 

%% file: 03methodology.tex
\section{Methodology}
\label{sec:methodology}

This study uses a mixed methods approach with a pragmatic philosophical world-view \cite{creswell2017research, creswell2021concise} to achieve a holistic understanding of what influences users' experience and their behaviour when they are managing their consent to cookies. A concurrent embedded strategy of inquiry was used, meaning that the qualitative and quantitative data were collected simultaneously. A thematic analysis was performed as part of the qualitative study \cite{content_thematic_2013}, while descriptive statistics and hypothesis tests were used as part of the quantitative analysis.

The participants took part in a usability test where they performed a set of tasks on five websites that exposed them to different cookie consent banners. The tests used both the participants' own computer and phone. After the usability test, the participants were asked to answer a survey. 

\subsection{Data collection}

Five websites were chosen to be included in the study. They were chosen based on popularity and to exhibit a diversity of deceptive patterns. The \emph{Tranco list} was used in order to favour the most popular websites \cite{tranco_article, tranco_list_KJYLW}. An overview of the websites can be found in Table~\ref{tab:website_overview}, while the details of the selection and which deceptive patterns that were present can be found in Appendix~\ref{app:website_selection}.

\input{tables/website_overview}

A total of 20 participants took part in the study. These participants were selected through convenience sampling. The sample size was chosen according to \textcite{20users}, which states that 20 participants is an optimal number to balance quantitative and qualitative measures.

\subsection{Usability test setup}

The tests were designed to measure how the participants interacted with cookie consent banners. The participants were asked to perform a set of given tasks interacting with the five websites on their own personal computer and mobile device. The test used the participants' own devices so that it was more likely that they would respond as they would by their own. In order to minimise the Hawthorne effect \cite{hawthorne}, the tasks were constructed in a way that made them interact with a cookie consent banner without knowing that it was this interaction that was measured.

A private browsing session was used in order to make sure that the cookie consent banners had not been answered beforehand, which was explained to the participants as being necessary in order to prevent autocompletion of searches. The order of websites and devices was randomised. Participants who had accepted the collection of cookies to at least one website were asked to remove their consent. The specific tasks are described in Appendix~\ref{app:test_procedure}.

The tests lasted for about 20 to 30 minutes. The usability test was conducted in person, where one author functioned as the facilitator. The \emph{think-aloud} protocol was utilised \cite{think_aloud_1980} and the tests were moderated. The time spent interacting with the cookie consent banners, their responses to them and how long it took to withdraw consent was measured by the facilitator.

\subsection{Survey}

The participants were asked to answer a survey consisting of 13 questions following the completion of the tasks. The themes of the questions included prior knowledge and understanding of cookies, feelings and attitudes and demographics. The specific questions in the survey can be found in Appendix~\ref{app:survey}.

After the survey, the participants were given a full explanation of the research aim of the study and how the collected data would be used. They were also given the justification for why this had not been explicitly disclosed to them before the usability tests, and were given the opportunity to ask questions about the study and to withdraw from the experiment if they wished to.

\subsection{Analysis}

Descriptive statistics and hypothesis tests were used to analyse the quantitative results. Non-parametric hypothesis tests were used to test if the expectation of the acceptance rate of the cookie consent banners and the times spent interacting with them were equal, in addition to testing if the expected time for giving and withdrawing consent differed. The code for the tests can be found at \url{https://github.com/LivHildeS/liv-hilde-master-thesis}.

As the intent of this article was to explore a wide range of possible connections related to interacting with cookie consent banners rather than arriving at conclusive evidence for a few specific claims, the hypothesis tests were not error corrected with respect to the number of tests performed. This is because doing so would incentivise performing fewer tests, which would contradict the aim of the article. Hence, the results should be interpreted carefully, in an exploratory way rather than a conclusive one.

\subsection{Ethical considerations}

The survey was constructed in order to minimise the collection of personal data, limited to vague age intervals and distinct groups for participants with or without IT expertise. The participants were informed that the data would be processed as part of this research and that it would be deleted afterwards.

In order to minimise bias from the Hawthorne effect, the participants were not informed about the main intent of the usability tests. The instructions given to the participants were true, but some details were withheld. The oral debriefing session aimed to preserve the participants' well-being, disclosing the aim of the study and the justification for withholding specific details \cite{deception}.

%% file: tables/website_overview.tex
\begin{table}[htbp]
    \centering
    \setlength{\tabcolsep}{5pt}
    \begin{tabular}{|r|r|r|}
        \hline
        \textbf{Website} & \textbf{\makecell{\textbf{Tranco} \\ \textbf{ranking}}} & \textbf{Owned by} \\
        \hline
        \texttt{google.com}    &       1    & Google LLC \\
        \texttt{facebook.com}  &       4    & Meta Platforms \\
        \texttt{finn.no}       &    1098    & Schibsted \\
        \texttt{dnb.no}        &   17563    & DNB \\
        \texttt{dagens.no}     &   48608    & Mediehuset Dagens \\
        
        \hline
        \end{tabular}
    \caption[Website overview]{\textbf{Overview of selected websites included in the usability tests.} The Tranco ranking measures how popular the websites are \cite{tranco_list_KJYLW}.}
    \label{tab:website_overview}
\end{table}

%% file: 04analysis_and_results.tex
\section{Analysis and results}
\label{sec:analysis}

\subsection{Quantitative analysis}
\label{sec:quantitative_analysis}

The options and choices made by the users when interacting with the cookie consent banners can be found in Figure~\ref{fig:banners_flowchart_answered}. Most of the granular cookie selections were never chosen. Only one answer was made for accepting a subset of the cookies, which was for \texttt{dagens.no}, and has been treated as an ``accept'' in the following analysis.

\input{figures/banners_flowchart_answered}

The total number of accepts for each website and device can be found in Table~\ref{tab:website_statistics_accepts}, while the average response times and standard deviations (SD) can be found in Table~\ref{tab:website_statistics_time}. The websites had substantially different numbers of accepts and response times. Particularly, \texttt{dagens.no} had more accepts and a much longer average response time than the other websites. There were 20 participants who interacted with 5 websites on 2 different devices, making a total of 200 responses to cookie consent banners. The total number of accepts was 59, thus, the average acceptance rate was 29.5\%.

\input{tables/analysis_tables/website_tests/website_statistics_accepts}

\input{tables/analysis_tables/website_tests/website_statistics_time}

The statistical significance of the differences was tested with Friedman tests \cite{friedman1937use} and can be found in Table~\ref{tab:friedman}. The null hypothesis is that all the websites have the same expected rate of consent or the same expected time spent interacting with the cookie consent banners, which can safely be rejected in all the tests. The tests are also statistically significant when testing the devices independently.

\input{tables/analysis_tables/website_tests/friedman}

Tests comparing the different device types, which did not show significant differences, can be found in Appendix~\ref{app:more_hypothesis_tests}. There are also tests for comparing the response to the websites pairwise against each other, which showed that \texttt{dagens.no} was a clear outlier.

Figure~\ref{fig:answer_times_line_plot_computer} and Figure~\ref{fig:answer_times_line_plot_phone} show time distributions given the response type for computer and phone, respectively. Apart from \texttt{dagens.no}, which has substantially longer response time for the participant who rejected consent, the other websites do not show any distinct pattern. The response times on the devices are fairly similar.

\input{figures/answer_times_line_plot_computer}

\input{figures/answer_times_line_plot_phone}

Finally, the time spent withdrawing consent compared to giving it was examined. The participants who gave consent to at least one website were asked to withdraw it for one to three websites, depending on time spent in the usability testing. Recall that the legal requirements state that withdrawing consent must be as easy as giving it \cite{GDPR, ekomloven}. 

The hypothesis can be tested by comparing the average withdrawal time per participant to the average time spent on giving consent to the same websites. This is a repeated measure context, and with pairwise data, one can use the Wilcoxon signed-rank test \cite{wilcoxon1992individual}. By averaging over all websites, the participants can be assumed to be independent of each other. Some participants always rejected consent and thus never withdrew, and those are excluded from this test.

\input{tables/analysis_tables/website_tests/withdrawal_statistics}

The average consent and withdrawal times can be found in Table~\ref{tab:withdrawal_statistics}. The hypothesis test was highly significant, with a test statistic of 0, meaning that all withdrawals took longer than all consents, and a p-value of 0.000061. On average, it took 21 times longer to withdraw consent than to give it. This is consistent across all websites and devices, although some websites have much higher withdrawal times than others. However, the number of withdrawals per website is low, so one should be hesitant to make any conclusive remarks about comparing the websites to each other. Even so, one can confidently reject the null hypothesis in favour of the alternative hypothesis, which states that the expected time spent withdrawing consent is significantly higher than the time spent giving it.

\subsection{Thematic analysis}
\label{sec:qualitative_analysis}

The thematic analysis was conducted following the six phases of \textcite{thematic_analysis_2006}. The familiarising phase was done by re-reading the notes and data collected and by making more notes. The thematic analysis was done by the same researcher collecting the data, so the content was already familiar.

The coding phase was done manually with an inductive approach, in order to allow for sufficient exploration \cite{kelle2004computer, thematic_analysis_2006}. Both semantic and latent coding were utilised. The initial coding gave 119 codes from the open ended questions and 25 codes from notes taken during the usability tests, which were grouped into potential themes. 

The review was done by evaluating the internal homogeneity and external heterogeneity for all potential themes, reviewing whether the codes within a theme were sufficiently similar to each other and if they were different enough from codes placed within other themes \cite{internal_homogeneity_external_heterogeneity}. The final theme maps can be found in Figure~\ref{fig:theme_maps}, while temporary maps and themes from the coding phase can be found in Appendix~\ref{app:thematic_analysis}.

\input{figures/theme_maps}

The essence of the themes and subthemes \cite{thematic_analysis_2006} is described as follows:

\begin{itemize}
    \item \textbf{Negative feelings towards cookie consent banners:} The feelings participants have towards cookie consent banners were frequently expressed as negative. These feelings can be divided into two types, feelings of strong dislike and feelings of anxiousness. This is illustrated by the quotes below.

    \begin{quote}
        \textit{Hate them, even though I would rather know what they are trying to do. Dealing with the bad ones, which make you jump through hoops to say ``no'', takes a significant amount of my day.}
    \end{quote}
    
    \begin{quote}
        \textit{I have very strong negative feelings about them.}
    \end{quote}

    \item \textbf{Cookie consent banners are a cognitive burden:} Several factors contributed to participants experiencing dealing with cookie consent banners as a cognitive burden. Participants expressed annoyance or irritation with the banners and their manipulation attempts.

    \item \textbf{Users call out the manipulation in cookie consent banners:} Participants repeatedly pointed out the manipulative tactics and deceptiveness of cookie consent banners. For instance, they commented on specific design features such as the use of \emph{colour to nudge} users towards accepting cookies and \emph{making it hard to say no}. Additionally, they frequently referenced established terminology within the field of user experience design and behavioural psychology, like the terms \emph{nudging} or \emph{dark patterns}. The awareness of this deceptive nature is illustrated by the quote below:
    
    \begin{quote}
        \textit{They are often deceptive in nature, trying to make you accept them by being disingenuous about what they do while maybe having a miniature link explaining everything about how it really is. }
    \end{quote}

    This shows that despite falling for the deceptive patterns, the users are aware of them being used.

    \item \textbf{Conditional willingness to consent:} Though most participants expressed rejecting cookie consent as their general preference, several participants emphasised two conditions under which the willingness to consent to cookies would be affected and potentially increased. It was stated that a high \emph{perceived trustworthiness} of the service requesting consent or an expected \emph{value in return} for their consent could increase the chances of consenting.

    \begin{quote}
        \textit{Giving consent for me depends on the amount of trust I have for the website. I try to decline for all shady websites as much as possible.}
    \end{quote}
        
    \item \textbf{Appreciate being given a choice:} Despite the participants frequently expressing negative feelings towards the cookie consent banners, a positive aspect of the banners that was brought up was the ability to \emph{have a choice}.
    
    \item \textbf{Wanting privacy without friction:} Most of the participants stated that their general preference is to not be tracked, and showed a desire to have their privacy intact and under their control, but without requiring too much effort, as illustrated below.

    \begin{quote}
        \textit{I always chose the minimum consent I was able to give in a short amount of time, scanning through the options available to me.}
    \end{quote}

    \item \textbf{Varying determination to reject:} Participants' responses demonstrated a varying determination to reject cookie consent, even though the general preference for most participants was to remain private and share less information. The observed rejection strategy can be divided into the categories \emph{uncompromising rejection}, \emph{steady rejection} and \emph{attempted rejection}.

    \item \textbf{Users are made to consent:} Several participants reported consenting in spite of it being against their preferences, as a result of being cornered. This was manifested in various ways, as participants frequently pointed to the amount of effort required to reject as a factor nudging them towards consent, leading to them consenting as a result of being \emph{worn down}. Participants also commented on the fact that access to the actual \emph{service was withheld until a response was given}, and the desire to move forward and proceed with their tasks pushed them to give a quick response, without necessarily paying attention to what their response was. Others expressed feeling \emph{forced to consent}, as options to reject were hidden or downplayed in the user interface.

\end{itemize}

An elaboration of the discussion around the themes, including more quotes from the participants supporting these claims, can be found in Appendix~\ref{app:themes}.

%% file: figures/banners_flowchart_answered.tex
\begin{figure}[htbp]
    \centering
    \includegraphics[width=0.99\linewidth]{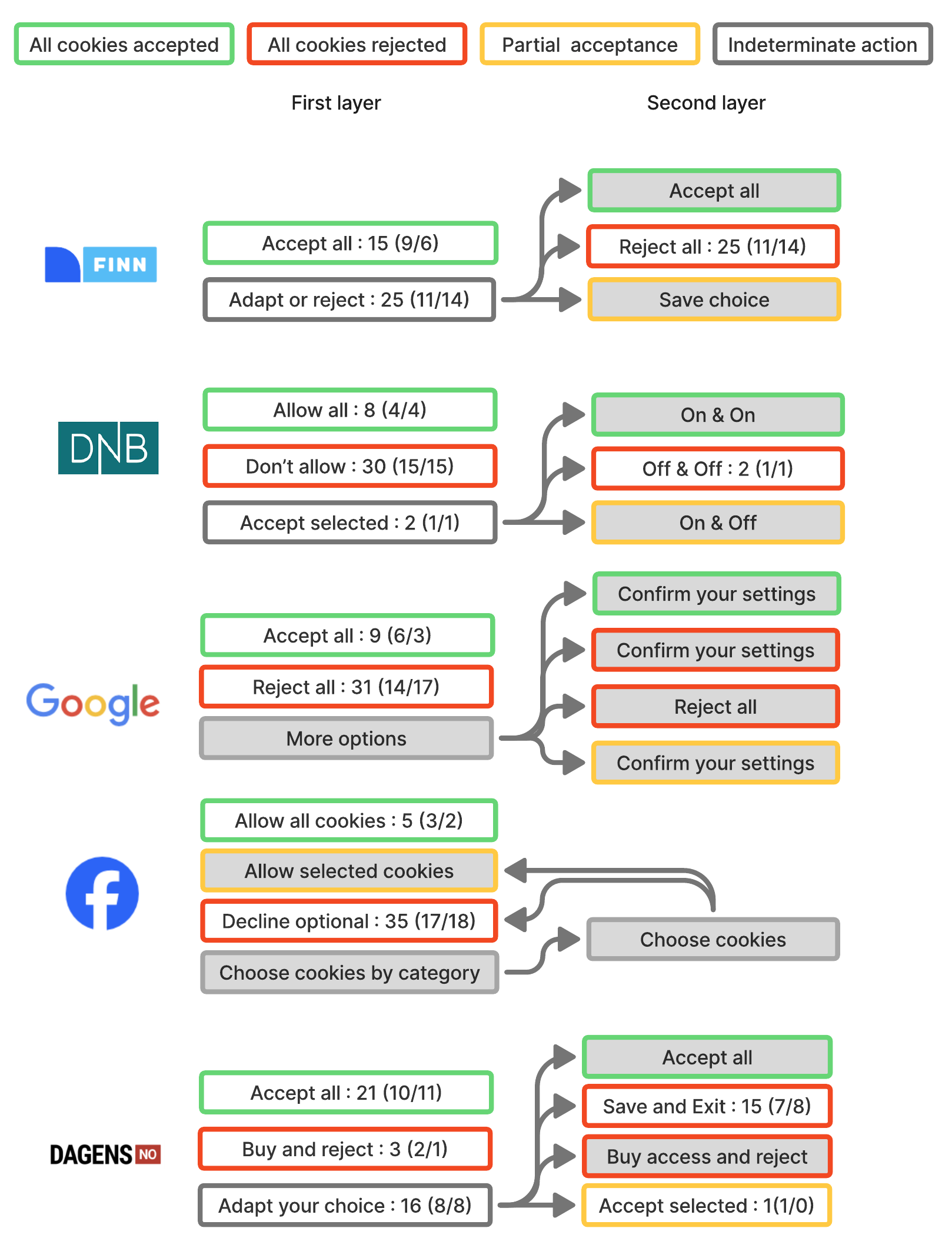}
    \caption[Flowchart with answers]{\textbf{Flowchart highlighting the choices made by the users.} The numbers indicate the total number of selections, selections on computer and selections on phone (option : total (computer / phone)). Options that were never selected have gray background colours.}
    \label{fig:banners_flowchart_answered}
\end{figure}

%% file: tables/analysis_tables/website_tests/website_statistics_accepts.tex
\begin{table}[htbp]
    \centering
    \begin{tabular}{|l|r|r|r|}
        \hline
        \textbf{Website} & \textbf{Computer} & \textbf{Phone} & \textbf{Total} \\
        \hline
        Facebook & 3 & 2 & 5 \\
        DNB & 4 & 4 & 8 \\
        Google & 6 & 3 & 9 \\
        Finn & 9 & 6 & 15 \\
        Dagens & 11 & 11 & 22 \\
        Sum & 33 & 26 & 59 \\
        \hline
    \end{tabular}
    \caption{ \textbf{Total number of cookie consent banner accepts} for the different websites on both devices. There were 20 participants. }
    \label{tab:website_statistics_accepts}
\end{table}

%% file: tables/analysis_tables/website_tests/website_statistics_time.tex
\begin{table}[htbp]
    \centering
    \footnotesize
    \begin{tabular}{|l|@{\hspace{4pt}}r@{\hspace{4pt}}|@{\hspace{4pt}}r@{\hspace{4pt}}|@{\hspace{4pt}}r@{\hspace{4pt}}|}
        \hline
        \textbf{Website} & \textbf{Computer} & \textbf{Phone} & \textbf{Total} \\
         & \textbf{Mean (SD)} & \textbf{Mean (SD)} & \textbf{Mean (SD)} \\
        \hline
        Facebook & 4.75 (2.17) & 4.15 (2.03) & 8.9 (3.81) \\
        DNB & 5.25 (4.12) & 4.2 (3.89) & 9.45 (5.57) \\
        Google & 4.15 (2.85) & 3.3 (1.34) & 7.45 (3.15) \\
        Finn & 4.95 (2.72) & 4.25 (1.68) & 9.2 (3.98) \\
        Dagens & 19.9 (26.19) & 8.3 (7.00) & 28.2 (26.49) \\
        Average & 7.8 (13.26) & 4.84 (4.14) & 12.64 (14.49) \\
        \hline
    \end{tabular}
    \caption{\textbf{Average time spent on cookie banners} for the different websites given in seconds.}
    \label{tab:website_statistics_time}
\end{table}

%% file: tables/analysis_tables/website_tests/friedman.tex
\begin{table}[htbp]
    \centering
    \footnotesize
    \begin{tabular}{|l|l|@{\hspace{4pt}}r@{\hspace{4pt}}|@{\hspace{4pt}}r@{\hspace{4pt}}|}
        \hline
        \textbf{Test variable} & \textbf{Device} & \textbf{Statistic} & \textbf{p-value} \\
        \hline
        Consent accepts & Computer & 16.741 & \textbf{0.00217046} \\
        Consent accepts & Phone & 19.538 & \textbf{0.00061583} \\
        Consent accepts & Both & 20.409 & \textbf{0.00041468} \\
        Answer time & Computer & 32.742 & \textbf{0.00000135} \\
        Answer time & Phone & 15.172 & \textbf{0.00435666} \\
        Answer time & Both & 37.825 & \textbf{0.00000012} \\
        \hline
    \end{tabular}
    \caption{ \textbf{Consent acceptance rates and answer times are significantly different.} The websites were tested against each other with Friedman hypothesis tests.}
    \label{tab:friedman}
\end{table}

%% file: figures/answer_times_line_plot_computer.tex
\begin{figure}[htbp]
  \centering
  \includegraphics[width=0.99\linewidth]{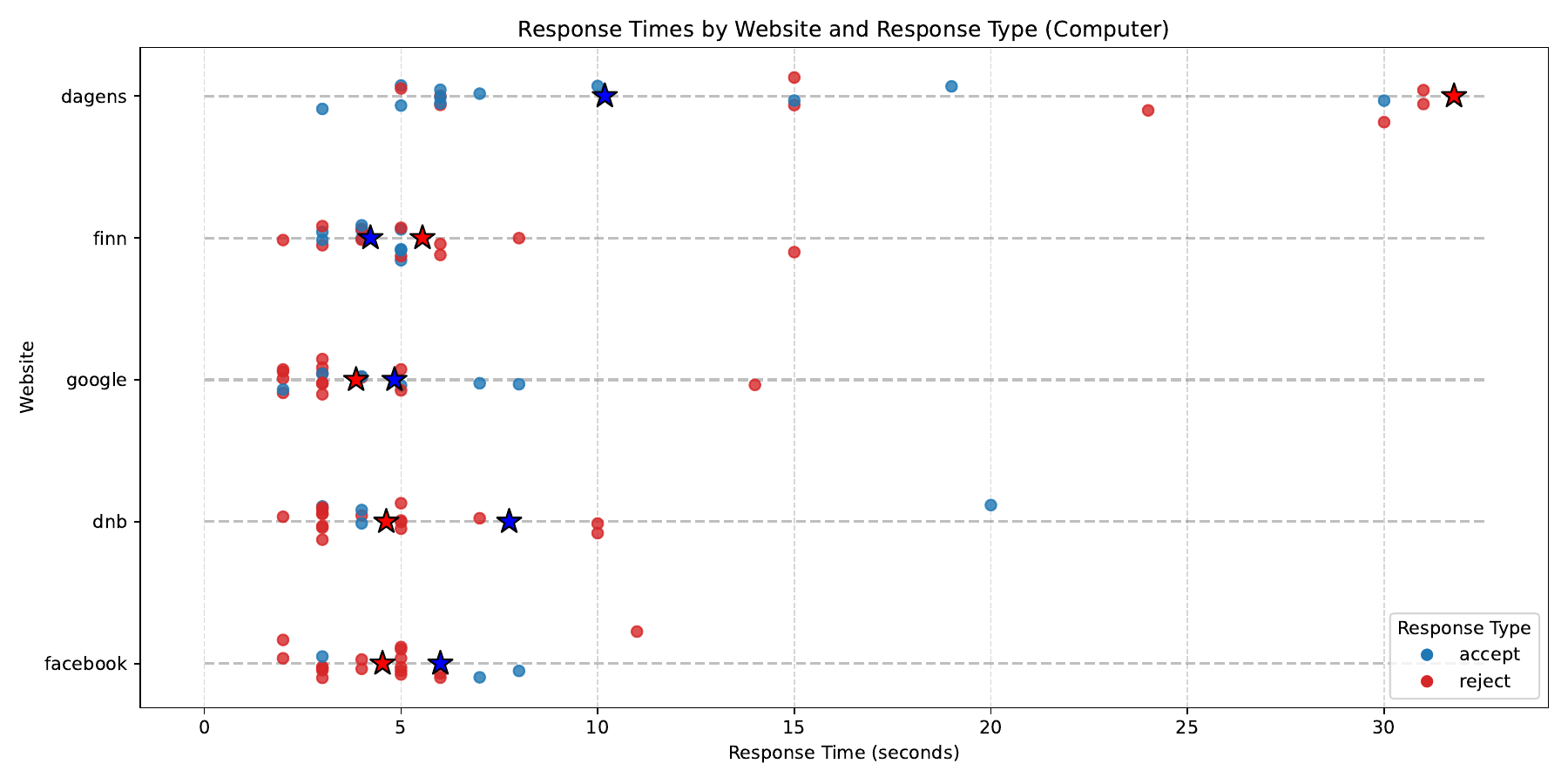}
  \caption[Answer time line plot computer]{\textbf{Cookie banner response times for the different websites on computer.} The stars mark the average response time for the accept and reject groups. The times are capped at 31 seconds, as some responses took much more time. Some random vertical spacing has been added to the dots to avoid too much overlap.}
  \label{fig:answer_times_line_plot_computer}
\end{figure}

%% file: figures/answer_times_line_plot_phone.tex
\begin{figure}[htbp]
  \centering
  \includegraphics[width=0.99\linewidth]{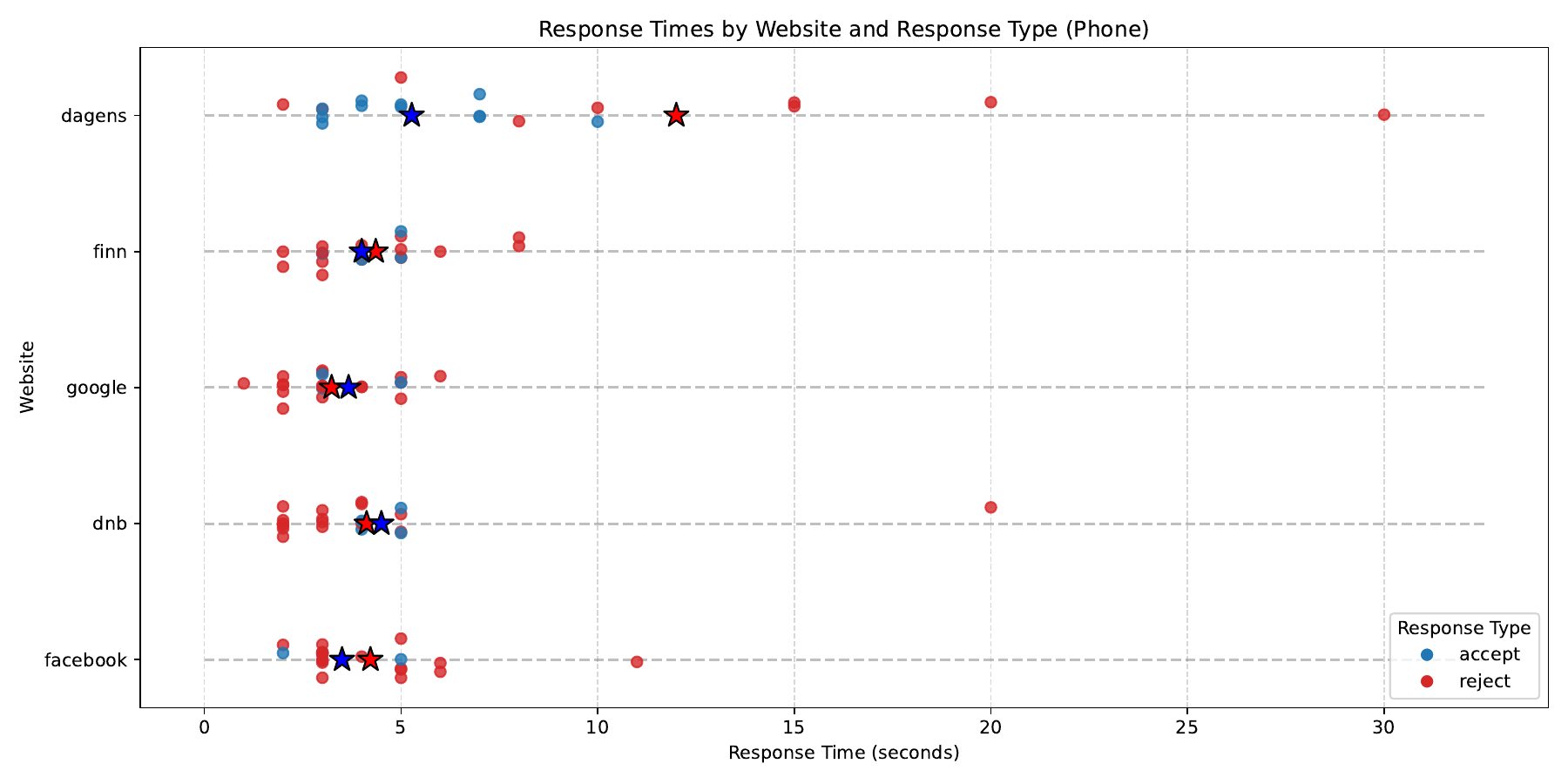}
  \caption[Answer time line plot phone]{\textbf{Cookie banner response times for the different websites on phone.} }
  %The star marks the average response time for the accept and reject groups. The times are capped at 31 seconds, as some responses to much more time. Some random vertical spacing has been added to the dots to avoid too much overlap.}
  \label{fig:answer_times_line_plot_phone}
\end{figure}

%% file: tables/analysis_tables/website_tests/withdrawal_statistics.tex
\begin{table*}[htbp]
    \centering
    \footnotesize
    \begin{tabular}{|l|l|r|r|r|r|}
        \hline
        \textbf{Website} & \textbf{Device} & \makecell{\textbf{Average} \\ \textbf{consent time}}  & \makecell{\textbf{Average} \\ \textbf{withdrawal time}} & \makecell{\textbf{Number of} \\ \textbf{consents}} & \makecell{\textbf{Number of} \\ \textbf{withdraws}} \\
        \hline
        Facebook & Computer & 6.00 & 180.00 & 3 & 1 \\
        Facebook & Phone & 3.50 & - & 2 & 0 \\
        Facebook & Both devices & 4.75 & 90.00 & 5 & 1 \\
        \hline
        DNB & Computer & 7.75 & 135.00 & 4 & 2 \\
        DNB & Phone & 4.50 & 220.00 & 4 & 1 \\
        DNB & Both devices & 6.12 & 177.50 & 8 & 3 \\
        \hline
        Google & Computer & 4.83 & 105.00 & 6 & 1 \\
        Google & Phone & 3.67 & - & 3 & 0 \\
        Google & Both devices & 4.25 & 52.50 & 9 & 1 \\
        \hline
        Finn & Computer & 4.22 & 133.75 & 9 & 4 \\
        Finn & Phone & 4.00 & 216.00 & 6 & 4 \\
        Finn & Both devices & 4.11 & 174.88 & 15 & 8 \\
        \hline
        Dagens & Computer & 10.18 & 70.00 & 11 & 2 \\
        Dagens & Phone & 5.27 & 88.33 & 11 & 3 \\
        Dagens & Both devices & 7.73 & 79.17 & 22 & 5 \\
        \hline
        Total & Computer & 6.60 & 124.75 & 33 & 10 \\
        Total & Phone & 4.19 & 104.87 & 26 & 8 \\
        Total & Both devices & 5.39 & 114.81 & 59 & 18 \\
        \hline
    \end{tabular}
    \caption{ \textbf{Withdrawing consent takes over 20 times longer than giving it.} The table shows average time spent giving consent and withdrawing it in seconds, and the number of participants who have given consent and withdrawn it. Comparing the average withdrawal time and corresponding consent time per participant gave highly significant results with a Wilcoxon signed-rank test, with a test statistic of 0 and \textbf{p-value of 0.000061}, rejecting the null hypothesis of equal consent and withdrawal times. }
    \label{tab:withdrawal_statistics}
\end{table*}

%% file: figures/theme_maps.tex
\begin{figure}[htbp]
    \centering
    \includegraphics[width=1\linewidth]{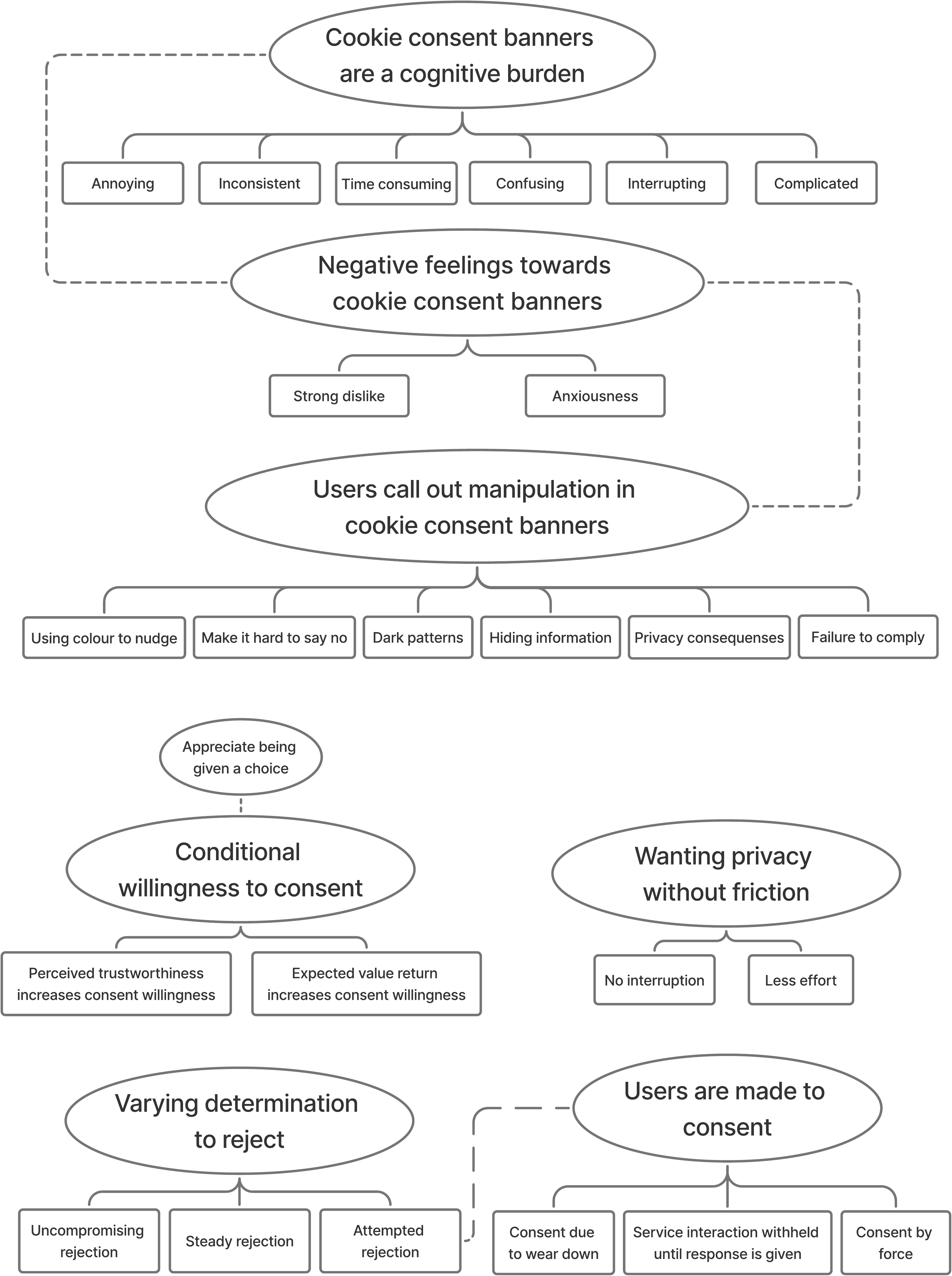}
    \caption[Theme maps]{\textbf{Finalised theme maps.} The figure illustrates the themes and subthemes identified during review phase of the thematic analysis, with the themes shown as ellipses and subthemes shown as rectangles. The dotted lines represent relations between themes while the solid lines illustrate a relation between themes and subthemes. }
    \label{fig:theme_maps}
\end{figure}

%% file: 05discussion.tex
\section{Discussion}
\label{sec:discussion}

\subsection{Merging and interpreting the results}

The thematic analysis revealed that the participants generally want privacy, but without having to work too hard in order to achieve it. The website \texttt{dagens.no} had the highest number of cookie consent banner accepts, which had a banner where the ``reject'' option was presented as ``Buy access and reject'', even though clicking this did not result in a payment request. The participants expressed dislike and annoyance of this banner, describing it as ``confusing'', ``time consuming'' and ``distrustful''. This could therefore be explained by the ``consent by force'' subtheme illustrated in Figure~\ref{fig:theme_maps}. Moreover, the website had the highest average time to respond to the cookie consent banners, about three times longer than the average. This suggests that the cookie consent banner was designed such that rejecting the cookies was troublesome, making the users accept even though they generally would not like to do so.

The participants had a different reaction to the second most accepted website, \texttt{finn.no}. Several participants expressed that they felt that the website was trustworthy and that accepting cookies could add value to the user experience. This is consistent with the website scoring highest on loyalty and customer satisfaction in the Norwegian customer barometer \cite{norsk_kundebarometer} and previous work on users' perceptions of cookie disclaimers \cite{this_website_uses_cookies}. The time to answer the cookie banner was not significantly different from the other websites. This suggests a ``conditional willingness to accept'', resulting in the website being accepted more than most of the other websites.

One of the clearest results from the quantitative analysis was that it took substantially longer to withdraw consent than to give it. According to GDPR Article 7, withdrawing consent shall be as easy as giving it \cite{GDPR}. This was not the case for any of the websites of this study, as the average time spent on withdrawing consent was over 20 times longer than giving it, consistent across all websites and devices.

\subsection{Answering the research questions}

\paragraph{RQ1 How do users respond to various cookie consent banners?}

The total number of cookie accepts was 59 out of 200 responses, making the acceptance rate of cookie consent 29.5\%. The website with the highest number of accepts was also the website with the strongest presence of deceptive patterns, namely \texttt{dagens.no}, with 22 accepts out of 40 responses. Participants repeatedly expressed annoyance from this website, and pointed out the fact that they felt forced to consent as a result of the ``consent or pay'' design in the cookie consent banner.

However, the second most accepted website was \texttt{finn.no} with 15 accepts out of 40 responses, which was perceived as trustworthy by the participants. By consenting to cookies, some participants expected that it would add value to their user experience. This demonstrates how drastically different aspects, namely deceptive patterns, trustworthiness and expected improvement to the customer experience, can influence the users' decision making and response to cookie consent banners.

\paragraph{RQ2 What are users' general privacy preferences and how do they compare to their responses to the cookie consent banners?}

To find users' general privacy preferences, questions \textbf{Q1, Q6} and \textbf{Q7} from the survey were used, displayed in Table~\ref{tab:nettskjema_report}. 

In response to ``\textbf{Q1} Privacy is about your right, as far as possible, to decide for yourself over your own personal data. To what extent are you concerned about privacy?'', 30\%  answered that they were ``very concerned'', while 40\% answered that they were ``quite concerned''  and 30\% answered that they were ``slightly concerned''. 
The responses to ``\textbf{Q6} Which statement best describes how you feel about sharing your information through cookies?''  showed that 90\% of the participants wanted as little information as possible about them and their online activity to be saved and shared. 
In response to ``\textbf{Q7} How do you usually respond to cookie consent banners?'', 60\% of the participants reported that they try to decline, but accept if rejecting takes too much effort, while 30\% of the participants answered that they actively pursue withholding consent.  

Answers to \textbf{Q1} and \textbf{Q6} indicate that privacy matters to the users, and that they generally prefer to share as little data as possible through cookies. However, the answers to \textbf{Q7} show that users are self-aware that they may not act according to these preferences, particularly if doing so requires effort. The acceptance rate of 29.5\% is aligned with the participants' self-reported consent behaviour from \textbf{Q7}, but does not align with their stated privacy and cookie consent preferences, which would suggest a lower acceptance rate. This gap may be explained by the influence of deceptive patterns and the effort required to reject cookies, which may cause users to respond in a way that is not in line with their actual preferences. 

While users' general preferences and level of privacy concern do influence how they manage cookie consent, their behaviour is also influenced by the amount of effort required, leading them to sometimes consent to cookies despite their general preferences. This insight further emphasises how users may be particularly vulnerable to the influence of deceptive patterns when managing consent.

\paragraph{RQ3 Is withdrawing consent as easy as giving it?}

This study found that withdrawing cookie consent took on average more than 20 times longer than giving it, as shown in Table~\ref{tab:withdrawal_statistics}. Therefore, the answer to \textbf{RQ3} is \emph{no}, withdrawing consent takes much more time, and is therefore not as easy. This is not in line with the requirement from GDPR \cite{GDPR}, and signifies a high cost of effort for users who want to manage their consent by withdrawing it.

\paragraph{RQ4 What are users’ general feelings, attitudes and preferences with regards to cookie consent banners?}

It was discovered in the thematic analysis that users may be willing to consent to cookies if conditions of trust and expected value return are met. While users appreciated having a choice, they generally had negative feelings towards cookie consent banners and experienced interacting with them as a cognitive burden. Users frequently called out the manipulative strategies and deceptive patterns that were used in the cookie consent banners. This demonstrates a level of awareness and critical thinking, despite that they still fell victim to manipulation and frequently were made to consent through service blocking, wear down or from feeling forced. Generally, users want to preserve their privacy and exercise their autonomy in a fair environment without interruption from other tasks or spending too much effort. In other words, \emph{users want privacy without friction}.

%% file: 06conclusions_and_future_work.tex
\section{Concluding remarks}
\label{sec:concluding_remarks}

\subsection{Conclusion}

This study investigated the gap between users' privacy preferences and their actual behaviour when they are managing their cookie consent. A usability test with 20 participants was conducted, where the participants engaged with five different websites' cookie consent banners on both computer and phone. In order to make the tests more realistic, the participants used their own devices and they were not informed prior to the test that their responses to cookie consent banners were of any importance. They were told that using private tabs was important to not get auto-suggestions to the task, while the real reason was that it made sure that the cookie consent banners showed up. Data about their responses to the cookie consent banners, time spent responding and their verbal utterances while doing so were recorded. The tests were followed up by a survey and debrief.

The \emph{privacy paradox} states that despite users reporting that they care about their privacy, they do not behave in a manner that preserves it \cite{norberg2007privacy}, but this article showed that the design of cookie consent banners substantially influenced the participants' consent management against their will. The participants reported that they generally preferred to reject consent to cookies, but in a frictionless way. The website with the highest number of accepts to the cookie consent banner had a high presence of deceptive patterns that annoyed the users, while the website with the second highest acceptance rate was generally considered trustworthy. Participants reported that they were willing to give a website their data if they trusted the service or it gave them value in terms of a better user experience, while some reported that they accepted cookies when the effort to reject was too high.

The results also revealed that withdrawing consent took on average more than 20 times longer than giving it. GDPR \cite{GDPR} states that withdrawing consent should be as easy as giving it. However, this did not seem to be the case for this study, suggesting that the current design of cookie consent does not adhere to this legal requirement.

This evidence suggests that the current form of acquiring consent through cookie banners is not ideal. Users want the opportunity to decide for themselves when to reject or consent to cookies, but the current solution with cookie consent banners is seen as a cognitive burden and nudges users towards consenting. Although there are laws regulating how the banners should be designed, they are not strictly adhered to, for instance by making it easier to accept than reject and by making it harder to withdraw consent than to give it. Despite the fact that users were aware of several manipulation strategies used, deceptive patterns often made users act against their preferences. This suggests that many cookie consent banner designs are effectively pursuing surface level compliance over real user agency and autonomy, leading to users consenting to cookies as a result of being pressured. This raises the question of whether consent is freely given as required by the GDPR \cite{GDPR}, and therefore if it is valid, since users are \emph{consenting by design}.

\subsection{Future work}

In light of the results in this article, a question arises concerning whether cookie banners are a good way to tackle the issue of consent. This article showed how the design of cookie consent banners substantially influenced users' consent management. Websites can influence their users' consent behaviours by intentionally designing the cookie banners with deceptive patterns, which the GDPR and other legislations try to mitigate \cite{GDPR}. It is therefore reasonable to pursue different ways of collecting consent. Future research should explore other approaches and strategies for collecting consent from website users in a way that does not disrupt the users' workflow nor nudge them towards consenting against their actual preferences. 

As this study utilised convenience sampling, the results are not directly generalisable to the broader population. Future studies could be conducted in order to validate and extend these findings. This would ideally be with an increased sample size, randomised selection and stratification ensuring the representation of marginalised groups and with multiple researchers performing the experiments and thematic coding.

%% file: 07limitations.tex
\section{Limitations}

Due to the convenience sampling, the results will not necessarily generalise to the rest of the population. The sample size of N=20 is high for a usability test, but not for the quantitative analysis. The study should mostly be interpreted as exploration around human interaction with cookie consent banners rather than conclusive evidence. As such, multiple hypothesis tests were carried out, without directly correcting the p-values with respect to the number of tests. This was done to not limit exploration, but means that p-values, especially those close to the border of significance, should be interpreted with care. While measures were taken for intra-rater reliability in the thematic analysis, there was only one person performing it, possibly limiting inter-rater reliability \cite{mcdonald2019reliability} and therefore reproducibility.

The experiments were set up to limit bias from the Hawthorne effect by focusing the tasks on non-cookie related tasks and by having the participants use their own devices, but the effect may still have had some influence. Some of the participants reported in the debriefing session that they had a suspicion that the experiment was related to cookies while taking the test, and some knew in advance that private browsing sessions did not retain choices from the cookie banners. This may have resulted in different behaviour among some of the participants compared to how they would have acted without being observed.

%% file: 08appendix.tex
\appendix

\section{Website Selection}
\label{app:website_selection}

The websites included in the usability test were chosen based on two selection criteria, that they had a presence of diverse deceptive patterns and that they were representative. The first criterion was chosen to answer the research questions and the second to increase the generalisability of the results. The websites were chosen to be representative by using the Tranco list for website popularity, where ``KJYLW'' was the id for the specific instance used \cite{tranco_article, tranco_list_KJYLW}, and by including \texttt{.no} domains as the study was conducted in Norway.

Both \texttt{google.com} and \texttt{facebook.com} were chosen directly due to their high popularity ranking, namely 1st and 4th. The websites ranked 2nd and 3rd were \texttt{microsoft.com} and \texttt{mail.ru}, which were not included since \texttt{microsoft.com} was not considered to be commonly used in casual daily interactions, and \texttt{mail.ru} was not considered to be relevant as it mainly contains content in Russian. 

The highest ranking \texttt{.no} domain on the Tranco list was \texttt{finn.no}, ranked at 1098 overall. The website has for several years been representing one of the leading companies in Norway on customer satisfaction and loyalty, and was ranked number one in the 2024 Norwegian customer barometer \cite{norsk_kundebarometer}. It was therefore included as it was both popular and possibly likely to inspire a certain level of trust in the participants, which made it an interesting addition to the study.  
Finally, \texttt{dnb.no} and \texttt{dagens.no} were chosen in order to represent a diversity of deceptive patterns while being relatively popular. The website \texttt{dnb.no} had a moderate presence of deceptive patterns, while \texttt{dagens.no} had an overwhelmingly manipulative design. Among the \texttt{.no} domains, \texttt{dnb.no} was ranked 19th and \texttt{dagens.no} 54th on the Tranco list.

Table~\ref{tab:website_pattern_matrix} gives an overview of the identified deceptive patterns in the selected websites' cookie consent banners, where the deceptive elements are specified and tied to a type of deceptive pattern. The deceptive patterns are described below.

\input{tables/website_pattern_matrix}

\begin{itemize}
    \item \textbf{Interface interference} demonstrates certain aspects of the interface being promoted over others \cite{dark_patterns_ux_2018}. This was found in the websites by having granular consent options less visible than accept buttons and visually elevating the accept button, as seen in Figure~\ref{fig:interface_inference_dagens}.
    
    \item \textbf{Positive framing} is when the service provider uses positive words to describe the option that benefits them or downplays the negative impacts that may follow for the user \cite{decieved_by_design_forbruker}. This pattern was observed in all of the cookie consent banners, by emphasising how enabling cookies would increase the quality of the service, for instance with ``provide personal tips'' or including ``relevant advertisements''.

    \item \textbf{Obstruction} means using obstacles or hindrances for complicating the path for completing a task, which leads to the user having to spend more time or clicks to achieve what they want, which is often used for choices that do not benefit the service provider \cite{obstruction}. This was demonstrated in several banners which required more clicks and steps for declining consent or choosing a granular option.

    \item \textbf{Forced action} is when the service provider presents an optional action as something necessary, which can be used to make the user choose something that is disadvantageous to them \cite{forced_action}. All cookie consent banners blocked the view of the websites until they were answered, even though the strictly necessary cookies do not require prior consent from a legal standpoint \cite{2002eprivacy}. The banner for \texttt{dagens.no}, as seen in Figure~\ref{fig:interface_inference_dagens}, presented the acceptance of cookies as the only free option to use the website. However, clicking ``Buy access and reject'' closed the banner and never resulted in a payment request.
    
    \item \textbf{Preselection} refers to one or more options being selected by default, nudging users towards selecting these options through the default effect bias \cite{defaults_bosch2016, pre_selection}. This deceptive pattern was implemented in the \texttt{finn.no} cookie consent banner when pursuing the granular consent option, as seen in Figure~\ref{fig:preselection_finn}. 
    
    \item \textbf{Trick wording} is when the use of misleading or confusing language tricks the user to perform an action \cite{trick_wording}. This was found in the \texttt{dagens.no} cookie consent banner as shown in Figure~\ref{fig:interface_inference_dagens}, as the ``Buy access and reject'' option does not require a payment.
    
\end{itemize}

\input{figures/interface_inference_dagens}
\input{figures/preselection_finn}

Note that some changes have been implemented since this study was conducted, but they remained consistent while conducting the study between January and February 2025. 

\section{Usability tests}
\label{app:usability_tests}

\subsection{Test procedure details}
\label{app:test_procedure}

All data collection sessions were conducted in a private room with one participant and the facilitator present. At the beginning of the session the participants were given an information letter that contained information about the session, tasks, duration and information about data management. After reading the information letter they were asked if they had any questions, which were answered if they had any. Once the participants were ready, they were given the paper with the set of tasks to perform, described in Table~\ref{tab:tasks_overview}. The participants then performed the tasks on both devices while the facilitator took notes.  

\input{tables/tasks_overview}

Once the usability testing was done, the participants were asked to answer the survey, which was located on the facilitator's computer. The facilitator remained in another part of the room whilst the participant answered the survey, so the participants were able to ask questions if they had any, but the facilitator could not see what the participants answered. Once the survey was answered, a conversation with a debriefing session was held.

\subsection{Pilot testing}
\label{app:pilot_testing}

Pilot testing on three different users was done in order to improve the usability tests and the survey. Both feedback from the users and observations made during the testing were used to make small updates to the tasks and setup. This resulted in some added complexity to the task instructions, limiting the number of withdrawals performed and revising the wording of the questions in the survey.

\section{Quantitative analysis details}
\label{app:analysis}

\subsection{Survey results}
\label{app:survey}

The questions, possible answers and results from multiple choice questions in the survey are presented in Table~\ref{tab:nettskjema_report}. Question 3 (\textbf{Q3}) asked seven true or false questions about cookies while \textbf{Q5} was a free-text field asking the users to elaborate on how they feel about cookie consent banners. \textbf{Q3} was aimed to be used as a measure of how much the participants knew about HTTP cookies, but as all the participants got a full score or close to it, it was not used in the analysis. \textbf{Q5} asked ``How do you feel about cookie consent banners?'' which was used in the thematic analysis.

The responses from questions \textbf{Q1} and \textbf{Q6} show that most participants are concerned about privacy, and that 90\% prefer to share as little information as possible through cookies. Almost all the participants state in \textbf{Q2} that they know what cookies are, and most say that they understand to at least some extent what kind of information they collect in \textbf{Q4}. In \textbf{Q7}, 60\% of the participants state that they usually try to decline cookies, but accept them if rejecting requires too much effort, whilst 30\% actively pursue withholding their consent. From \textbf{Q8} and \textbf{Q9} one can see that most participants have not previously withdrawn their consent to cookies, and most also do not know that withdrawing consent should be as easy as giving consent. \textbf{Q11} and \textbf{Q12} show that most of the participants are within the age range 20-39 years, and that 3/4 have an IT related background.

\input{tables/analysis_tables/overviews/nettskjema_report}

The answers to the survey were used to perform hypothesis tests to examine the rate of acceptance to cookie consent banners and the time spent interacting with them. The groups were constructed in the following way, where \textbf{Q2} and \textbf{Q6} were excluded due to being too unbalanced.

\begin{itemize}
    \item \textbf{Q1} was grouped by participants who were ``quite'' or ``very'' concerned about privacy and participants who were only ``slightly'' concerned.
    \item \textbf{Q4} was divided into participants who understand permissions to collect user data to ``some'' or a ``great'' extent, and the rest of the participants in the other group.
    \item \textbf{Q7} was grouped by participants who ``actively take steps to withhold my consent'', against the other participants.
    \item \textbf{Q8} was divided into participants who have withdrawn consent and those who had not or did not know.
    \item \textbf{Q9} was divided into participants who knew that withdrawing should be as easy as giving it, against the participants who did not know.
    \item \textbf{Q11} was grouped by participants who were younger than 30 years and those who were 30 years and older.
    \item \textbf{Q12} was divided based on participants having IT-related background or not.
\end{itemize}

Ideally, the groups would have been decided before investigating the data, in order to limit bias from having knowledge of the distribution of the data. However, due to the sample size, it was decided to group the participants after the surveys were answered in order to achieve more balanced groups.

The test performed was bootstrap confidence intervals (CI) \cite{1996bootstrap_ci}, as the data was not assumed to approximately follow some underlying distribution. The groups were tested with the total number of accepts and the time spent interacting with the cookie consent banners on both devices. The results can be found in Table~\ref{tab:bootstrap_main}, which shows that apart from participants who are concerned about privacy accepting less and using more time interacting with the cookie consent banners, most results are not significant.

\input{tables/analysis_tables/group_tests/boostrap_main}

\subsection{More hypothesis tests}
\label{app:more_hypothesis_tests}

The websites were tested pairwise against each other with pairwise Wilcoxon signed-ranked tests \cite{wilcoxon1992individual}, with the results in Table~\ref{tab:wilcoxon_total_accepts} and Table~\ref{tab:wilcoxon_total_times}. The website \texttt{dagens.no} is an outlier, having both significantly higher acceptance rates and answer times than most of the other websites, while most of the other websites do not have significant differences.

\input{tables/analysis_tables/website_tests/wilcoxon_total_accepts}

\input{tables/analysis_tables/website_tests/wilcoxon_total_time}

Hypothesis tests for whether the participants acted differently given the different devices used were also performed. Given a website, it was tested if the participants were more likely to give consent on computer or phone and if they spent longer time answering the cookie banner on computer or phone. As this is a pairwise repeated measures scenario, the Wilcoxon signed-ranked test \cite{wilcoxon1992individual} was used.

\input{tables/analysis_tables/website_tests/website_devices_tests_accepts}

\input{tables/analysis_tables/website_tests/website_devices_tests_time}

The results from the test can be found in Table~\ref{tab:website_devices_tests_accepts} and Table~\ref{tab:website_devices_tests_time}. The p-values are all high, except for the time spent on all websites aggregated. Thus, the devices did not significantly influence how the participant reacted. This is not surprising when inspecting the total number of accepts and average time spent per website in Table~\ref{tab:website_statistics_accepts} and Table~\ref{tab:website_statistics_time}, since the values are close. However, there is a slight tendency to give fewer accepts and spend shorter time when using phone, so the results may have become significant if a larger sample size was used.

\section{Thematic analysis details}

The themes from phase three of the thematic analysis can be found in Table~\ref{fig:theme_table1} and Table~\ref{fig:theme_table2}. No themes were removed in this process.

\label{app:thematic_analysis}

\input{figures/theme_table1}

\input{figures/theme_table2}

\input{figures/themes_partial_behaviour}

\input{figures/themes_partial_negative_feelings}

The themes were changed significantly in the fourth phase. For instance, the ``General consent behaviour'' theme was dissolved entirely, and became the themes ``Conditional willingness to consent'', ``Varying determination to reject'' and ``Users are made to consent''. Figure~\ref{fig:themes_partial_behaviour} shows how the old subthemes of ``General consent behaviour'' shown in Figure~\ref{fig:theme_table2} were transformed into the new themes and subthemes. Figure~\ref{fig:themes_partial_negative_feelings} shows how the ``What do they explicitly feel?'' theme was transformed into the new theme ``Negative feelings towards cookie consent banners''. Figure~\ref{fig:themes_partial_privacy_friction} shows how the ``What do they want?'' theme was transformed into the new ``Wanting privacy  without friction'' theme, and finally Figure~\ref{fig:themes_partial_consent_banners} shows how ``Cookie consent banners are'' was transformed into the ``Users call out the manipulation in cookie consent banners'', ``Cookie consent banners are a cognitive burden'' and ``Appreciate being given a choice'' themes. 

\input{figures/themes_partial_privacy_friction}

\input{figures/themes_partial_consent_banners}

\subsection{Deriving the themes}
\label{app:themes}

The following section provides an elaboration of the themes derived from the thematic analysis, along with quotes from the participants who supported the decisions. The quotes are taken from the verbal utterances the participants made during the usability test, \textbf{Q5} from the survey and in the debriefing session after the tests.

\textbf{Negative feelings towards cookie consent banners:} When participants were asked how they felt about cookie consent banners, the responses revealed a consistent feeling of \emph{strong dislike} towards the point of hate. Below are extracts from the responses given by two of the participants.  

\begin{quote}
    \textit{Hate them, even though I would rather know what they are trying to do. Dealing with the bad ones, which make you jump through hoops to say ``no'', takes a significant amount of my day.}
\end{quote}

\begin{quote}
    \textit{I have very strong negative feelings about them.}
\end{quote}

The extracts illustrate the active resentment the participants felt towards the cookie consent banners. The intensity of the language, such as the word ``\emph{hate}'' demonstrates the depth of these feelings.

In addition to the feelings of strong dislike, several participants expressed feelings of \emph{anxiousness} or worry with regards to cookie consent banners. Below are extracts from two participants responding to the question ``How do you feel about cookie consent banners?''. 

\begin{quote}
    \textit{I feel it is like someone trying to butt in to a conversation I am having.}
\end{quote}

\begin{quote}
    \textit{I fear that the cookies we accept today often are more verbose compared to how web-pages were structured 5 years ago.}
\end{quote}

\begin{quote}
    \textit{Every page nowadays uses google analytics, but how google actually stores the information or uses it internally is often hard to find.}
\end{quote}

These extracts illustrate the participants' feelings of anxiousness from being affected by cookies, in addition to feelings of concern about the long-term consequences of accepting them. This points to the fact that the feeling of anxiousness is not only tied to the consent situation, but also to the way that users' data is being used by data collectors.  

\textbf{Cookie consent banners are a cognitive burden:} The participants experienced dealing with cookie consent banners as a cognitive burden, and several factors seem to contribute to the mental load of this burden.

\begin{quote}
    \textit{They are annoying and try to trick users into accepting cookies.}
\end{quote}

The inconsistency of the banner designs makes them hard to adapt to, further adding to the mental load as shown in the following extracts:

\begin{quote}
    \textit{I feel that it is challenging that they come in so many different versions. It is easy to find ``not accept'' in some of them, and others are very complicated, and I don't find ``do not accept cookies''.}
\end{quote}

\begin{quote}
    \textit{Their appearance also varies a lot from site to site, making it hard to create a ``muscle memory'' of how to quickly answer them.}
\end{quote}

Several participants commented on the complexity of the banners, the time it takes to deal with them, and that they are being interrupted from their workflow as a burden.

\begin{quote}
    \textit{They slow me down most of the time.}
\end{quote}

\begin{quote}
    \textit{\texttt{Dagens} was a pain. Very confusing and hard to understand what I did and did not agree to.}
\end{quote}

\begin{quote}
    \textit{Breaking my workflow.}
\end{quote}

\begin{quote}
    \textit{I think it is frustrating when it's overcomplicated and too many choices.}
\end{quote}

Combined, these aspects of the cookie consent banners amount to a cognitive burden for the participants, increasing the mental load and having a negative impact on the user experience.   

\textbf{Users call out the manipulation in cookie consent banners:} In addition to participants experiencing the cognitive burdens of dealing with the cookie consent banners, they also repeatedly point out the manipulation tactics and deceptiveness of cookie consent banners.

\begin{quote}
    \textit{They seem to often be designed to make you accept them, either through colours or making it hard to reject cookies.}
\end{quote}

\begin{quote}
    \textit{I try to click on decline, sometimes I get tricked by dark patterns.}
\end{quote}

Others point out the general deceptiveness of the banners, discussing the potential \emph{privacy consequences} of accepting them, the fact that many banners \emph{fail to comply} with relevant regulations and requirements for valid consent and often obscure or \emph{hide information} about the data being collected and how this data is used, as demonstrated by the extracts below.

\begin{quote}
    \textit{I know in several cases, if you have enough cookies you are able in some cases to track down behaviour patterns, movement and other very private information about the user.}
\end{quote}

\begin{quote}
    \textit{They are often deceptive in nature, trying to make you accept them by being disingenuous about what they do while maybe having a miniature link explaining everything about how it really is. }
\end{quote}

\begin{quote}
    \textit{A lot of them do not follow EU regulation.}
\end{quote}

\begin{quote}
    \textit{Not something a user actually reads/can reasonably consent to.}
\end{quote}

The fact that several participants are pointing out these elements of manipulation and deceptiveness in the cookie consent banners demonstrates a level of awareness amongst users. This shows that although users may fall victim to this manipulation in practice, it does not necessarily imply a lack of critical thinking. Rather, users consciously recognize the ways in which their right to a valid consent environment is being undermined, giving context to the previously discussed theme ``Negative feelings towards cookie consent banners''. 

\textbf{Conditional willingness to consent:} Although the participants generally wanted to reject cookies, several participants emphasised that they were increasingly willing to accept if the website had a level of \emph{perceived trustworthiness} or that accepting cookies gave them some expected \emph{value in return}.

\begin{quote}
    \textit{Giving consent for me depends on the amount of trust I have for the website. I try to decline for all shady websites as much as possible.}
\end{quote}

\begin{quote}
    \textit{ \texttt{DNB} site is also a trustworthy known website, so consenting there was not an issue. The only one giving me pause was \texttt{Dagens}.}
\end{quote}

\begin{quote}
    \textit{I accepted the ones I trust, like \texttt{Finn} and \texttt{DNB}. I rejected the ones I don't trust.}
\end{quote}

\begin{quote}
    \textit{For trusted sites (like \texttt{DNB}) I would not be as sceptical. For example, \texttt{Dagens} which clearly focuses heavily on ads, I would be more sceptical to consent.}
\end{quote}

\begin{quote}
    \textit{ I am more comfortable giving consent to websites that I am sure will bring value to my internet browsing, like \texttt{Finn}.}
\end{quote}

In summary, these conditions point to the fact that users' general consent preference may not be constant, but rather that it is context dependent and that users' willingness to consent may be affected by an evaluation of the perceived risks and expected benefits.

\textbf{Appreciate being given a choice:} Despite the negative feelings towards cookie consent banners, several participants commented that they preferred \emph{having a choice}.

\begin{quote}
    \textit{Annoying, but better than not being able to choose.}
\end{quote}

\begin{quote}
    \textit{But for me, who rejects cookies, it is nice since I think I can reject them.}
\end{quote}

\begin{quote}
    \textit{I think they are annoying, but understand why they are there.}
\end{quote}

Although users are critical of cookie consent banners, they still value being able to consent. This points out an important nuance, that even though the cookie consent banners can be described as generally disliked by users, this does not mean that users want a situation where the banners are removed without a replacement solution ensuring their autonomy.

\textbf{Wanting privacy without friction:} Most of the participants state a desire to have their privacy intact and under their control, as demonstrated by the extracts below. 

\begin{quote}
    \textit{I do not want to be tracked, I do not want targeted advertising, I really just want to be left alone when trying to accomplish something.}
\end{quote}

\begin{quote}
    \textit{I always chose the minimum consent I was able to give in a short amount of time, scanning through the options available to me.}
\end{quote}

However, it also becomes clear that they want their privacy to remain intact without having to invest too much effort to keep it that way, and without being interrupted during activities unrelated to privacy management. In other words, they generally want to remain private and be able to manage their privacy and cookie consent, but with \emph{less effort} and \emph{without interruption} during other activities.
 
\textbf{Varying determination to reject:} Participants' responses demonstrate a varying determination to reject cookie consent. The determination to reject can be divided into categories. Some participants report exercising \emph{uncompromising rejection}, as shown by the extracts below. 

\begin{quote}
    \textit{Sometimes I decide to not view the content if it is displayed on the condition that I give consent.}
\end{quote}

\begin{quote}
    \textit{I often out of spite either choose to check everything that has an indication that I can choose, or do not use the page at all if I can help it.}
\end{quote}

Though this uncompromising rejection is practised, it does not mean that the same individual will consistently act in this way, and it might be a strategy that is used only under certain circumstances. In this sense, this theme is a relation to the ``Conditional willingness to consent'' theme.

In contrast to the uncompromising rejection, some participants report a more steady approach to rejection, clearly stating that they generally try to reject cookies, but with less intensity. The use of the word ``try'' indirectly indicates the presence of obstacles or friction preventing them from achieving what they want. This is shown by the participants frequently referring to the level of ``ease of rejecting'' as an important factor impacting whether or not they can respond with rejection successfully, as demonstrated by the extract below. 

\begin{quote}
    \textit{I don't want to allow cookies, but I also don't want to use too much effort when having to decline.}
\end{quote}

Considering this, two more varieties or strategies of determination to reject can be identified, namely the strategy of \emph{steady rejection}, which is less determined than uncompromising rejection, and the least determined \emph{attempted rejection} strategy.

\textbf{Users are made to consent:} Participants reported having consented in spite of it being against their preferences, due to being \emph{worn down}, or that the \emph{service was withheld until a response was given}, and the desire to move forward and to proceed with their tasks pushed them to give a quick response, without necessarily paying attention to what their response was. Others expressed feeling \emph{forced to consent}. These manifestations are demonstrated by the extracts below. 

\begin{quote}
    \textit{I accepted cookies when it was hard to find ``do not accept'', and it needed more work.}
\end{quote}

\begin{quote}
    \textit{Declined the easy ones, accepted the one where I had to pay to decline.}
\end{quote}

\begin{quote}
    \textit{\texttt{Dagens} seemed like they wanted me to purchase a subscription and the only other option was to accept. The others I was able to decline.}
\end{quote}

\begin{quote}
    \textit{They can sometimes cover the content I wanted to read.}
\end{quote}

\begin{quote}
    \textit{The easiest way to solve the task, to come forward.}
\end{quote}

This illustrates how design choices can undermine the validity of consent, by making it hard to reject and blocking interaction with the service until a response to the consent request is given.

%% file: tables/website_pattern_matrix.tex
\begin{table*}[htbp]
\centering
    \begin{tabular}{|l|c|c|c|c|c|}
        \hline
        \textbf{Dark Pattern} &
        \textbf{Dnb} &
        \textbf{Dagens} &
        \textbf{Facebook} &
        \textbf{Google} &
        \textbf{Finn} \\ \hline
        
        \shortstack[l]{\rule{0pt}{1.2em}\textbf{Interface interference:}\\Reject and accept \\ are not equal} &
        x & x & x &   & x \\ \hline
        
        \shortstack[l]{\rule{0pt}{1.2em}\textbf{Interface interference:}\\ Granular consent is \\ less visible than accept} &
        x & x & x & x & x \\ \hline
        
        \shortstack[l]{\rule{0pt}{1.2em}\textbf{Obstruction:}\\Rejecting requires \\unnecessary steps} &
          & x &   &   & x \\ \hline
        
        \shortstack[l]{\rule{0pt}{1.2em}\textbf{Obstruction:}\\Granular consent \\requires extra steps} &
        x & x & x & x & x \\ \hline
        
        \shortstack[l]{\rule{0pt}{1.2em}\textbf{Forced action:}\\Consent or pay} &
          & x &   &   &   \\ \hline
        
        \shortstack[l]{\rule{0pt}{1.2em}\textbf{Forced action:}\\Must respond to proceed} &
        x & x & x & x & x \\ \hline
        
        \shortstack[l]{\rule{0pt}{1.2em}\textbf{Positive framing:}\\Benefits emphasized,\\privacy impact downplayed} &
        x & x &   &   & x \\ \hline
        
        \shortstack[l]{\rule{0pt}{1.2em}\textbf{Preselection:}\\Response is preselected} &
          &   &   &   & x \\ \hline
        
        \shortstack[l]{\rule{0pt}{1.2em}\textbf{Trick wording:}\\Fake payment option} &
          & x &   &   &   \\ \hline

    \end{tabular}

\caption[Website matrix]{\textbf{Presence of dark patterns in cookie consent banners.}}
\label{tab:website_pattern_matrix}

\end{table*}

%% file: figures/interface_inference_dagens.tex
\begin{figure}[htbp]
  \centering
  \includegraphics[width=0.9\linewidth]{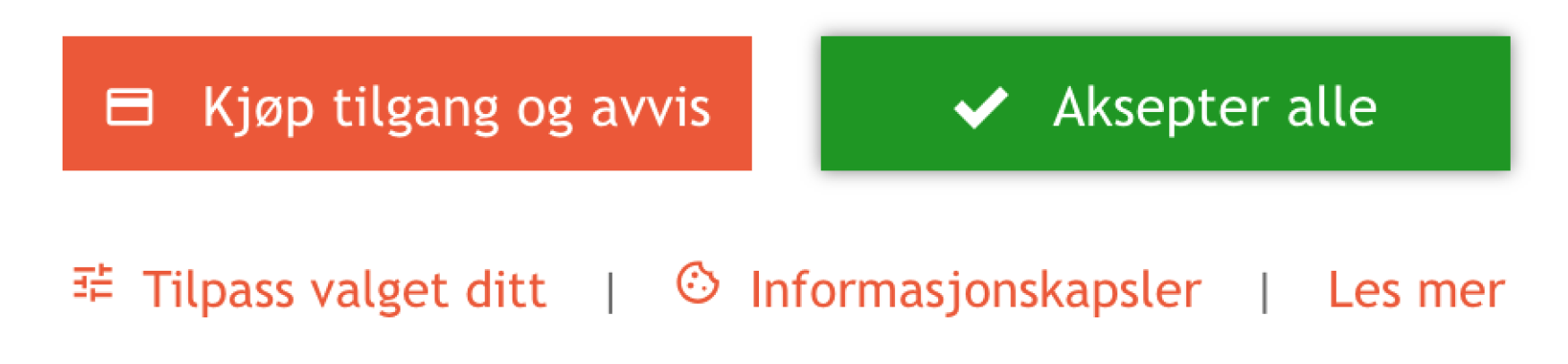}
  \caption[Dagens.no cookie banner]{\textbf{The \texttt{dagens.no} cookie consent banner showing multiple deceptive patterns.} Notice the elevation effect and colour of the accept button, in addition to the lack of comparable visibility for the granular consent option (``Tilpass valget ditt''). The user is presented with an option to ``Accept all'' or to ``Buy access and reject'', but choosing the latter did not result in a payment request.}
  \label{fig:interface_inference_dagens}
\end{figure}

%% file: figures/preselection_finn.tex
\begin{figure}[htbp]
    \centering
    \includegraphics[width=0.65\linewidth]{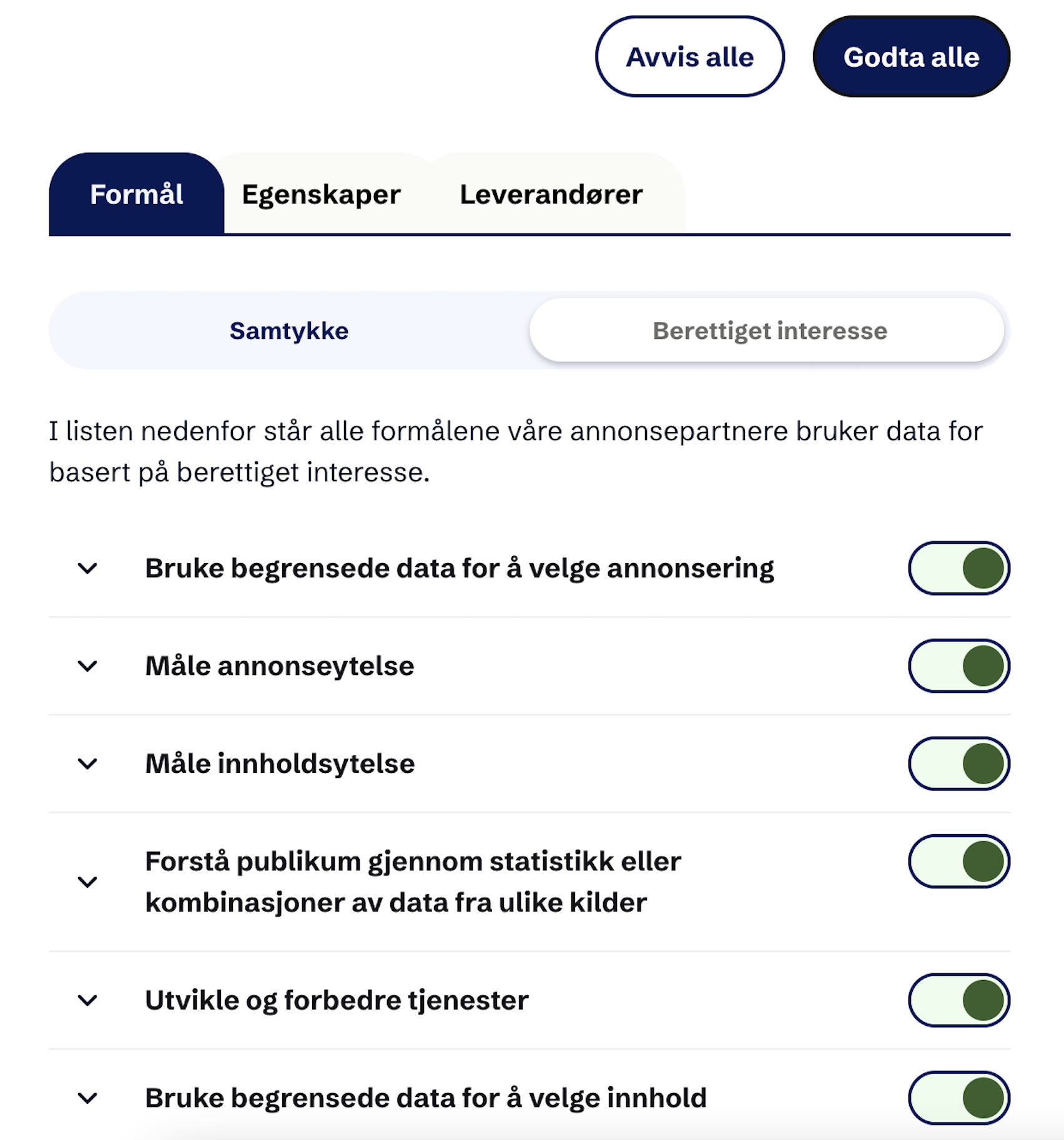}
    \caption[Finn.no preselection]{\textbf{Preselection pattern in the \texttt{finn.no} cookie consent banner}. The options for consent are already preselected as accepted.}
    \label{fig:preselection_finn}
\end{figure}

%% file: tables/tasks_overview.tex
\begin{table*}[htbp]
    \centering
    \begin{tabular}{|l|p{10cm}|}
        \hline
        \textbf{Website} & \textbf{Tasks} \\
        \hline
        \textbf{Finn} & 
        1. Open a new private tab. \newline
        2. Go to \texttt{finn.no}. \newline
        3. Go to ``Torget''. \newline
        4. Find a chair for sale in Østfold for under 200NOK. \newline
        5. Leave this tab standing, and open a new private tab. \\
        \hline
        \textbf{Facebook} & 
        1. Go to \texttt{facebook.com}. \newline
        2. Go to their ``help'' pages. \newline
        3. Find their ``reporting abuse'' pages. \newline
        4. Respond to their ``Was this helpful'' pop-over. \newline
        5. Leave this tab standing, and open a new private tab. \\
        \hline
        \textbf{Dagens} & 
        1. Go to \texttt{dagens.no}. \newline
        2. Find their search bar. \newline
        3. Find an article about ``Taco''. \newline
        4. Leave this tab standing, and open a new private tab. \\
        \hline
        \textbf{Google} & 
        1. Go to \texttt{google.com}. \newline
        2. Find a picture of Queen Sonja of Norway. \newline
        3. Find out the name of her mother. \newline
        4. Find a picture of her mother. \newline
        5. Leave this tab standing, and open a new private tab. \\
        \hline
        \textbf{DNB} & 
        1. Go to \texttt{dnb.no}. \newline
        2. Find their contact pages. \newline
        3. Start a conversation with their chatbot. \newline
        4. Find out what the age limit for a BSU-account is. \newline
        5. Leave this tab standing, and open a new private tab. \\
        \hline
    \end{tabular}
    \caption[Tasks overview]{\textbf{The tasks given to the participants in the usability tests}. The instructions were handed out on paper to the participants.}
    \label{tab:tasks_overview}
\end{table*}

%% file: tables/analysis_tables/overviews/nettskjema_report.tex
\begin{table*}[htbp]
    \centering
    \footnotesize
    \begin{tabular}{|p{12.5cm}|r|}
        \hline
        \textbf{Response} & \textbf{Count} \\
        \hline
        \textbf{Q1.} Privacy is about your right, as far as possible, to decide for yourself over your own personal data. To what extent are you concerned about privacy? & ~ \\[0.2em]
        Very concerned & 6 \\
        Quite concerned & 8 \\
        Slightly concerned & 6 \\
        Don't know & 0 \\
        \hline
        \textbf{Q2.} Do you know what cookies are? & ~ \\[0.2em]
        Yes & 19 \\
        No & 1 \\
        \hline
        \textbf{Q4.} Most websites ask for consent to collect information about you through cookies. To what extent would you say you understand what kind of information different websites request permission to collect? & ~ \\[0.2em]
        To a great extent & 2 \\
        To some extent & 11 \\
        Neither nor & 2 \\
        To a small extent & 3 \\
        Not at all & 2 \\
        Don't know & 0 \\
        \hline
        \textbf{Q6.} Which statement best describes how you feel about sharing your information through cookies? & ~ \\[0.2em]
        I want as little information as possible about me and my online activity to be saved and shared. & 18 \\
        I want information about me and my online activity to be saved and shared, since it improves my user experience. & 1 \\
        I am indifferent to my information and online activity being saved and shared. & 1 \\
        Don't know & 0 \\
        \hline
        \textbf{Q7.} How do you usually respond to cookie consent banners? & ~ \\[0.2em]
        I ignore them and leave them open. & 0 \\
        I choose the easiest option, whether it's accept or decline. & 1 \\
        I try to decline when possible, but accept if rejecting is too much effort. & 12 \\
        I actively take steps to withhold my consent. & 6 \\
        I consent because it improves my user experience. & 1 \\
        Don't know & 0 \\
        \hline
        \textbf{Q8.} Have you ever withdrawn your consent to cookies, after first having given it? & ~ \\[0.2em]
        Yes & 5 \\
        No & 14 \\
        Don't know & 1 \\
        \hline
        \textbf{Q9.} Were you aware that legally, withdrawing consent must be as easy as giving it? & ~ \\[0.2em]
        Yes & 7 \\
        No & 13 \\
        Don't know & 0 \\
        \hline
        \textbf{Q11.} How old are you? & ~ \\[0.2em]
        15 - 19 years & 0 \\
        20 - 29 years & 8 \\
        30 - 39 years & 7 \\
        40 - 49 years & 2 \\
        50 - 59 years & 2 \\
        60+ years & 1 \\
        \hline
        \textbf{Q12.} Do you have an IT-related background? & ~ \\[0.2em]
        Yes, programming and/or design related & 15 \\
        Yes, other & 0 \\
        No & 5 \\
        \hline
    \end{tabular}
    \caption{ \textbf{Answers to the quantitative questions in the survey.} }
    \label{tab:nettskjema_report}
\end{table*}

%% file: tables/analysis_tables/group_tests/boostrap_main.tex
\begin{table*}[htbp]
    \centering
    % \footnotesize
    \begin{tabular}{|l|r|r|r|}
        \hline
        \textbf{Group comparison} & \textbf{Mean diff} & \textbf{Cohen's d} & \textbf{95\% CI} \\
        \hline
        \multicolumn{4}{|l|}{\textbf{Total accepts on both devices}} \\
        \hline
        Quite or very concerned about privacy & -3.88 & -1.57 & \textbf{[-6.88, -1.10]} \\
        Understands cookie consent & 2.93 & 1.08 & \textbf{[0.11, 5.90]} \\
        Actively withholds consent & -0.78 & -0.26 & [-3.07, 1.70] \\
        Have withdrawn consent & -1.36 & -0.44 & [-3.74, 1.21] \\
        Aware of withdrawal ease & 0.52 & 0.17 & [-2.20, 3.54] \\
        Age & -2.00 & -0.68 & [-4.21, 0.04] \\
        IT background & -0.60 & -0.19 & [-4.20, 2.27] \\
        \hline
        \multicolumn{4}{|l|}{\textbf{Average banner answer time on both devices}} \\
        \hline
        Quite or very concerned about privacy & 2.41 & 0.78 & \textbf{[0.58, 4.50]} \\
        Understands cookie consent & -1.53 & -0.48 & [-3.93, 0.65] \\
        Actively withholds consent & 2.50 & 0.71 & [-0.96, 6.21] \\
        Have withdrawn consent & 3.01 & 1.02 & \textbf{[0.14, 6.41]} \\
        Aware of withdrawal ease & 0.80 & 0.25 & [-1.99, 3.88] \\
        Age & -0.18 & -0.05 & [-2.97, 3.03] \\
        IT background & -0.83 & -0.25 & [-5.25, 2.55] \\
        \hline
    \end{tabular}
    \caption{ \textbf{Bootstrap CIs and statistics for the survey groups}. The CIs not containing 0 reject the null hypothesis, and are marked in \textbf{bold}. The number of bootstrap samples used was 10000. }
    \label{tab:bootstrap_main}
\end{table*}

%% file: tables/analysis_tables/website_tests/wilcoxon_total_accepts.tex
\begin{table}[htbp]
    \centering
    \footnotesize
    \begin{tabular}{|l|l|r|r|r|}
        \hline
        \textbf{Website 1} & \textbf{Website 2} & \textbf{n} & \textbf{Statistic} & \textbf{p-value} \\
        \hline
        Facebook & DNB & 3 & 0.0 & 0.250000 \\
        Facebook & Google & 3 & 0.0 & 0.250000 \\
        Facebook & Finn & 8 & 0.0 & \textbf{0.007812} \\
        Facebook & Dagens & 10 & 0.0 & \textbf{0.001953} \\
        DNB & Google & 5 & 6.0 & 1.000000 \\
        DNB & Finn & 10 & 10.0 & 0.091797 \\
        DNB & Dagens & 12 & 7.0 & \textbf{0.010742} \\
        Google & Finn & 5 & 0.0 & 0.062500 \\
        Google & Dagens & 10 & 3.0 & \textbf{0.011719} \\
        Finn & Dagens & 11 & 16.0 & 0.127930 \\
        \hline
    \end{tabular}
    \caption{ \textbf{Pairwise Wilcoxon tests on the consent acceptance rates on both devices.} The website \texttt{dagens.no} has significantly different responses than the other websites. The number \textbf{n} marks the number of observations with different responses for each pair of website, as the Wilcoxon test only uses these. }
    \label{tab:wilcoxon_total_accepts}
\end{table}

%% file: tables/analysis_tables/website_tests/wilcoxon_total_time.tex
\begin{table}[htbp]
    \centering
    \footnotesize
    \begin{tabular}{|l|l|r|r|r|}
        \hline
        \textbf{Website 1} & \textbf{Website 2} & \textbf{n} & \textbf{Statistic} & \textbf{p-value} \\
        \hline
        Facebook & DNB & 17 & 70.5 & 0.775622 \\
        Facebook & Google & 18 & 42.5 & 0.059278 \\
        Facebook & Finn & 16 & 64.5 & 0.855977 \\
        Facebook & Dagens & 18 & 0.0 & \textbf{0.000194} \\
        DNB & Google & 17 & 55.0 & 0.307020 \\
        DNB & Finn & 20 & 84.5 & 0.438809 \\
        DNB & Dagens & 20 & 2.5 & \textbf{0.000127} \\
        Google & Finn & 18 & 53.5 & 0.160328 \\
        Google & Dagens & 19 & 0.0 & \textbf{0.000131} \\
        Finn & Dagens & 19 & 4.5 & \textbf{0.000268} \\
        \hline
    \end{tabular}
    \caption{ \textbf{Pairwise Wilcoxon tests on the banner response time on both devices.} Once again \texttt{dagens.no} is an outlier. }
    \label{tab:wilcoxon_total_times}
\end{table}

%% file: tables/analysis_tables/website_tests/website_devices_tests_accepts.tex
\begin{table}[htbp]
    \centering
    \begin{tabular}{|l|r|r|}
        \hline
        \textbf{Website} & \textbf{Statistic} & \textbf{p-value} \\
        \hline
        Facebook & 0.000 & 0.317311 \\
        DNB & 5.000 & 1.000000 \\
        Google & 0.000 & 0.083265 \\
        Finn & 8.000 & 0.256839 \\
        Dagens & 1.500 & 1.000000 \\
        All & 45.000 & 0.089555 \\
        \hline
    \end{tabular}
    \caption{ \textbf{Wilcoxon's signed rank test for whether accepts were given on the different devices.} }
    \label{tab:website_devices_tests_accepts}
\end{table}

%% file: tables/analysis_tables/website_tests/website_devices_tests_time.tex
\begin{table}[htbp]
    \centering
    \begin{tabular}{|l|r|r|}
        \hline
        \textbf{Website} & \textbf{Statistic} & \textbf{p-value} \\
        \hline
        Facebook & 36.000 & 0.157929 \\
        DNB & 45.500 & 0.138431 \\
        Google & 33.000 & 0.378689 \\
        Finn & 21.000 & 0.278762 \\
        Dagens & 53.000 & 0.090733 \\
        All & 915.000 & \textbf{0.006580} \\
        \hline
    \end{tabular}
    \caption{ \textbf{Wilcoxon's signed rank test for the times spent answering cookie banners on the different devices.} }
    \label{tab:website_devices_tests_time}
\end{table}

%% file: figures/theme_table1.tex
\begin{figure}[htbp]
    \centering
    \includegraphics[width=0.95\linewidth]{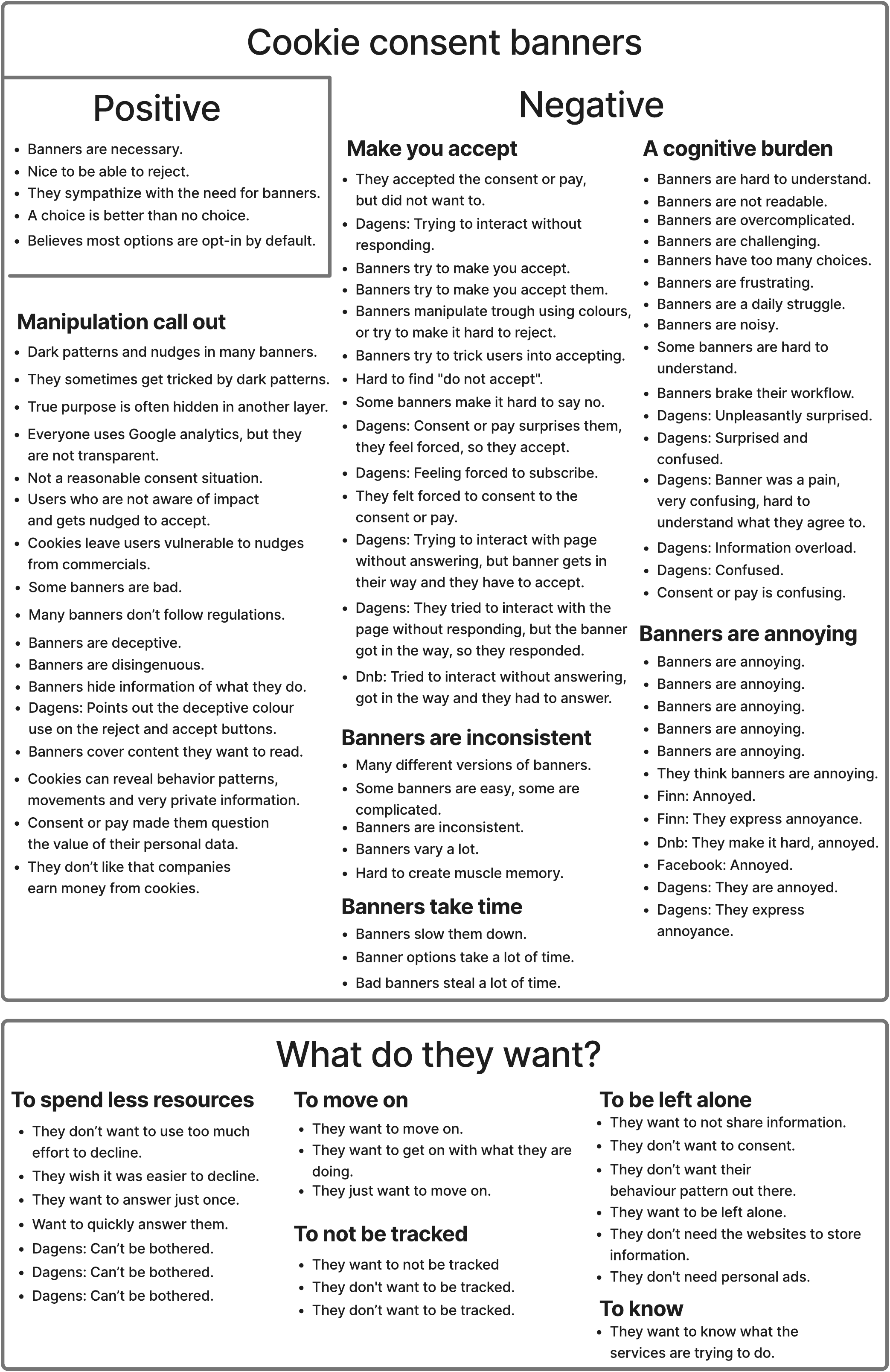}
    \caption[Theme table 1]{\textbf{Themes and subthemes with corresponding codes, part 1.} This was the product of phase 3 in the thematic analysis.}
    \label{fig:theme_table1}
\end{figure}

%% file: figures/theme_table2.tex
\begin{figure}[htbp]
    \centering
    \includegraphics[width=0.95\linewidth]{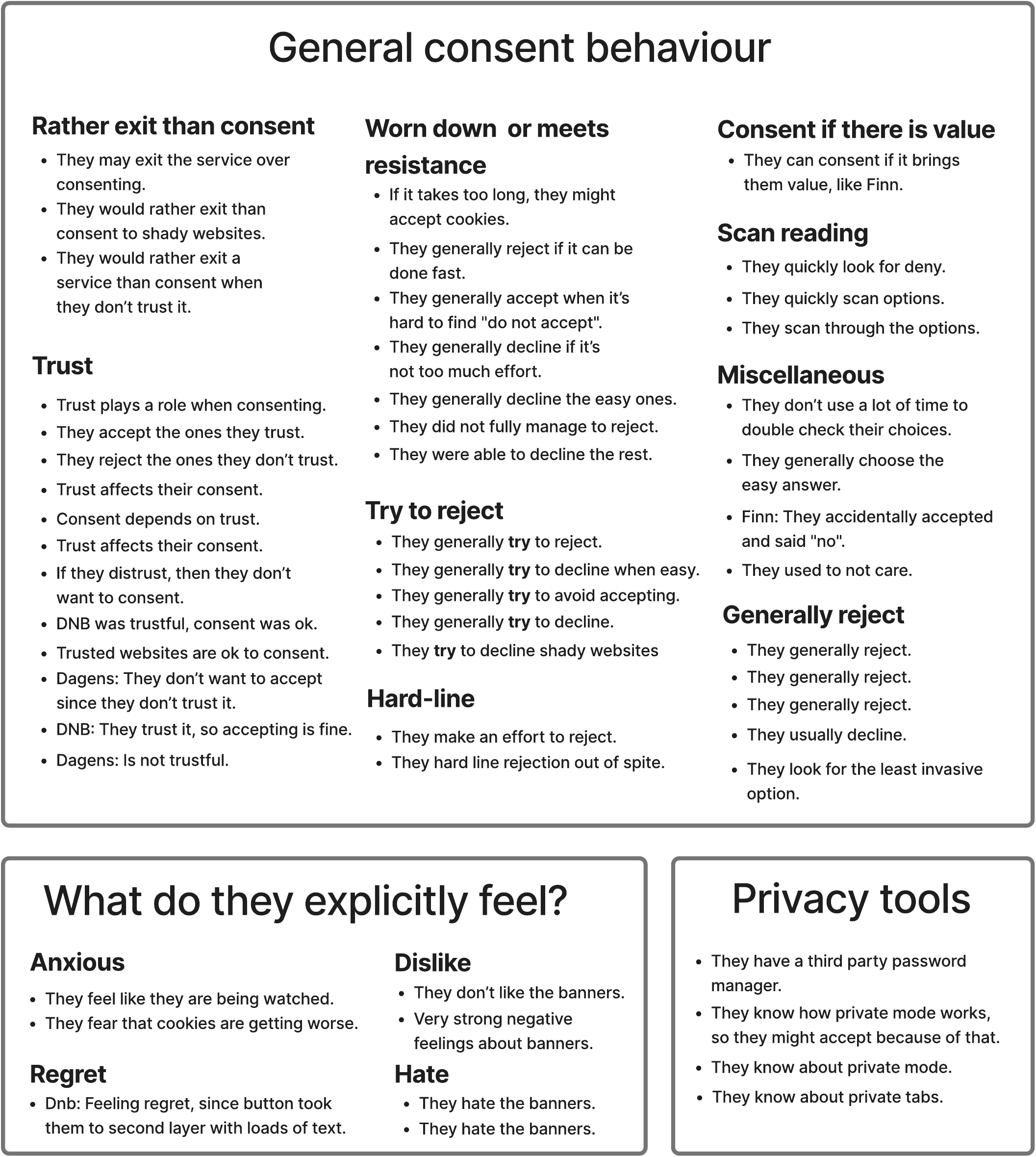}
    \caption[Theme table 2]{\textbf{Themes and subthemes with corresponding codes, part 2.} This was the product of phase 3 in the thematic analysis.}
    \label{fig:theme_table2}
\end{figure}

%% file: figures/themes_partial_behaviour.tex
\begin{figure}[htbp]
    \centering
    \includegraphics[width=1\linewidth]{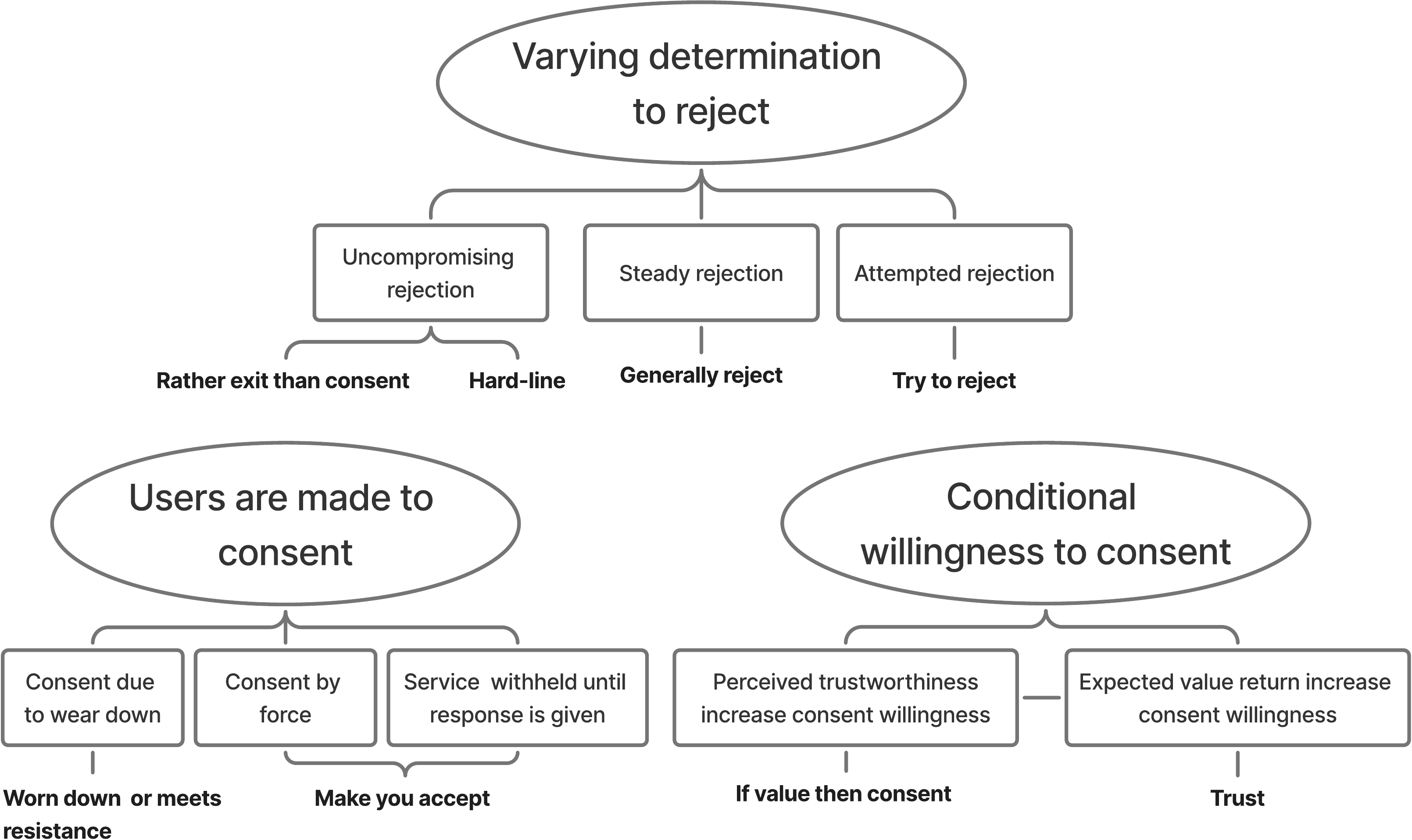}
    \caption[Themes partial cookie consent behaviour]{\textbf{Revised themes for ``General cookie consent behaviour''.} The figure shows how the theme ``general cookie consent behaviour'' from phase 3 was transformed into three revised themes used in phase 4.}
    \label{fig:themes_partial_behaviour}
\end{figure}

%% file: figures/themes_partial_negative_feelings.tex
\begin{figure}[htbp]
    \centering
    \includegraphics[width=0.5\linewidth]{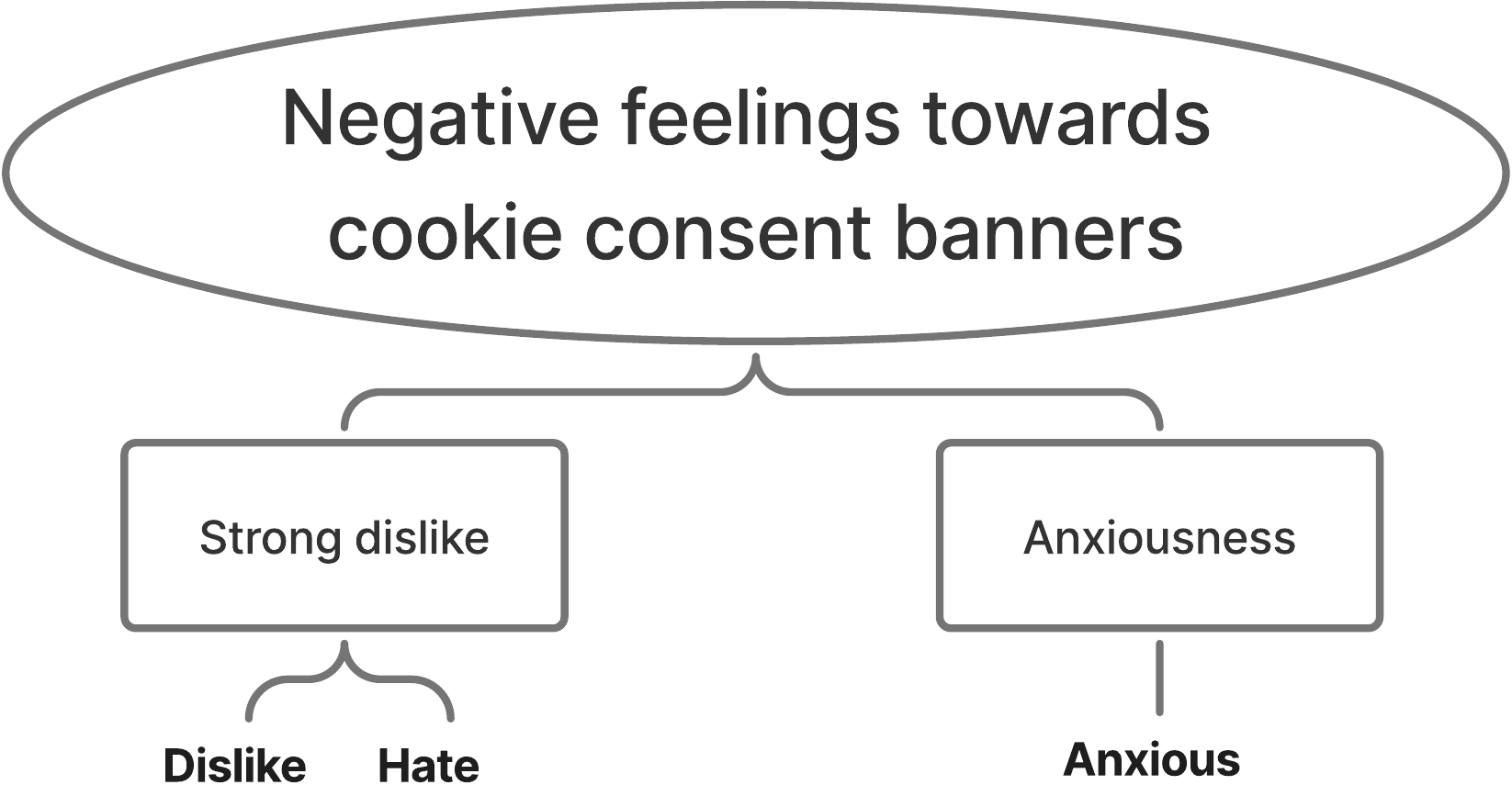}
    \caption[Themes partial negative feelings]{\textbf{Revised themes for ``How do they feel?''.} The figure shows how one theme from phase 3 was transformed in phase 4.}
    \label{fig:themes_partial_negative_feelings}
\end{figure}

%% file: figures/themes_partial_privacy_friction.tex
\begin{figure}[htbp]
    \centering
    \includegraphics[width=0.45\linewidth]{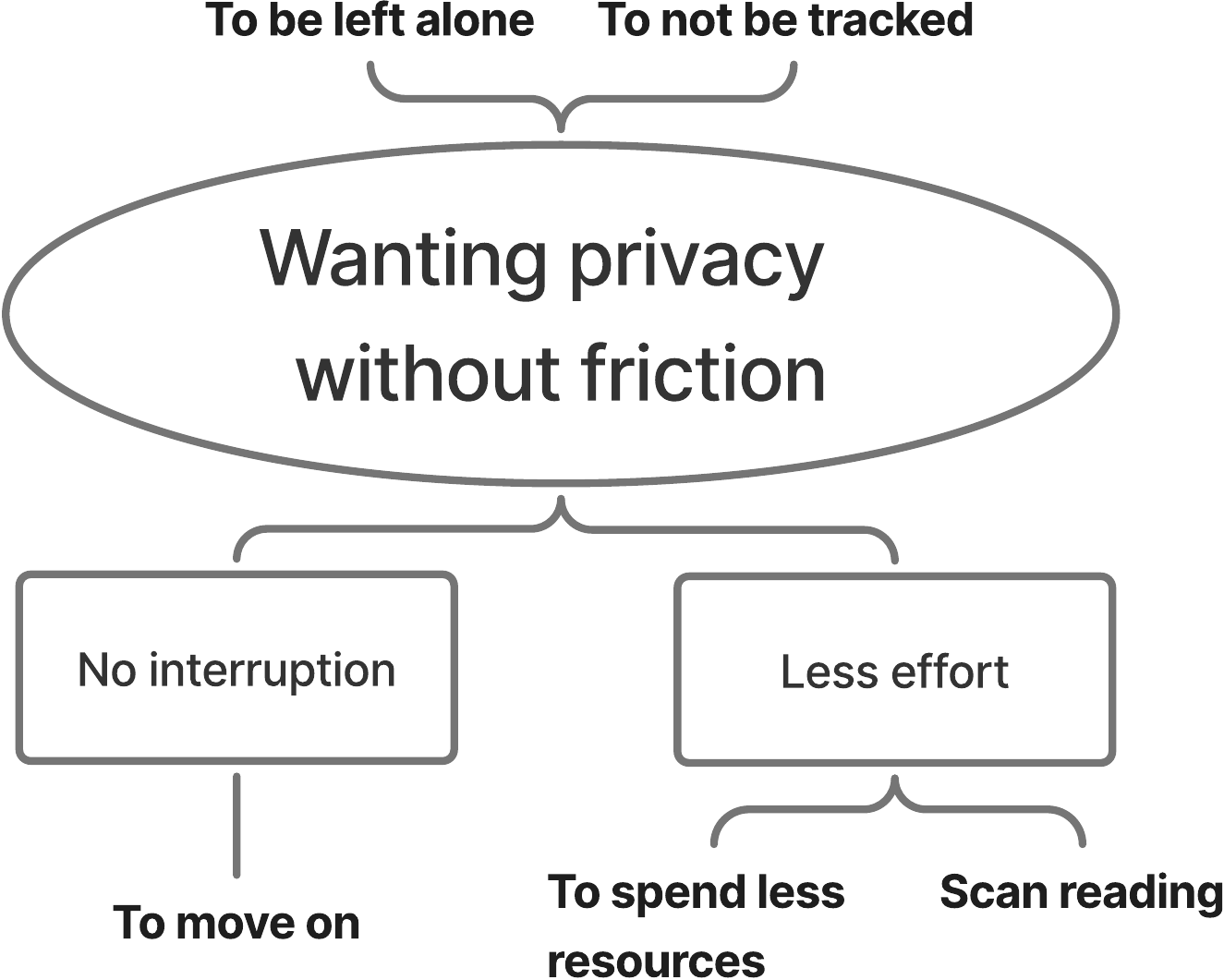}
    \caption[Themes partial privacy friction]{\textbf{Revised themes for ``What do they want?''.} The figure shows how a theme from phase 3 was transformed in phase 4.}
    \label{fig:themes_partial_privacy_friction}
\end{figure}

%% file: figures/themes_partial_consent_banners.tex
\begin{figure}[htbp]
    \centering
    \includegraphics[width=1\linewidth]{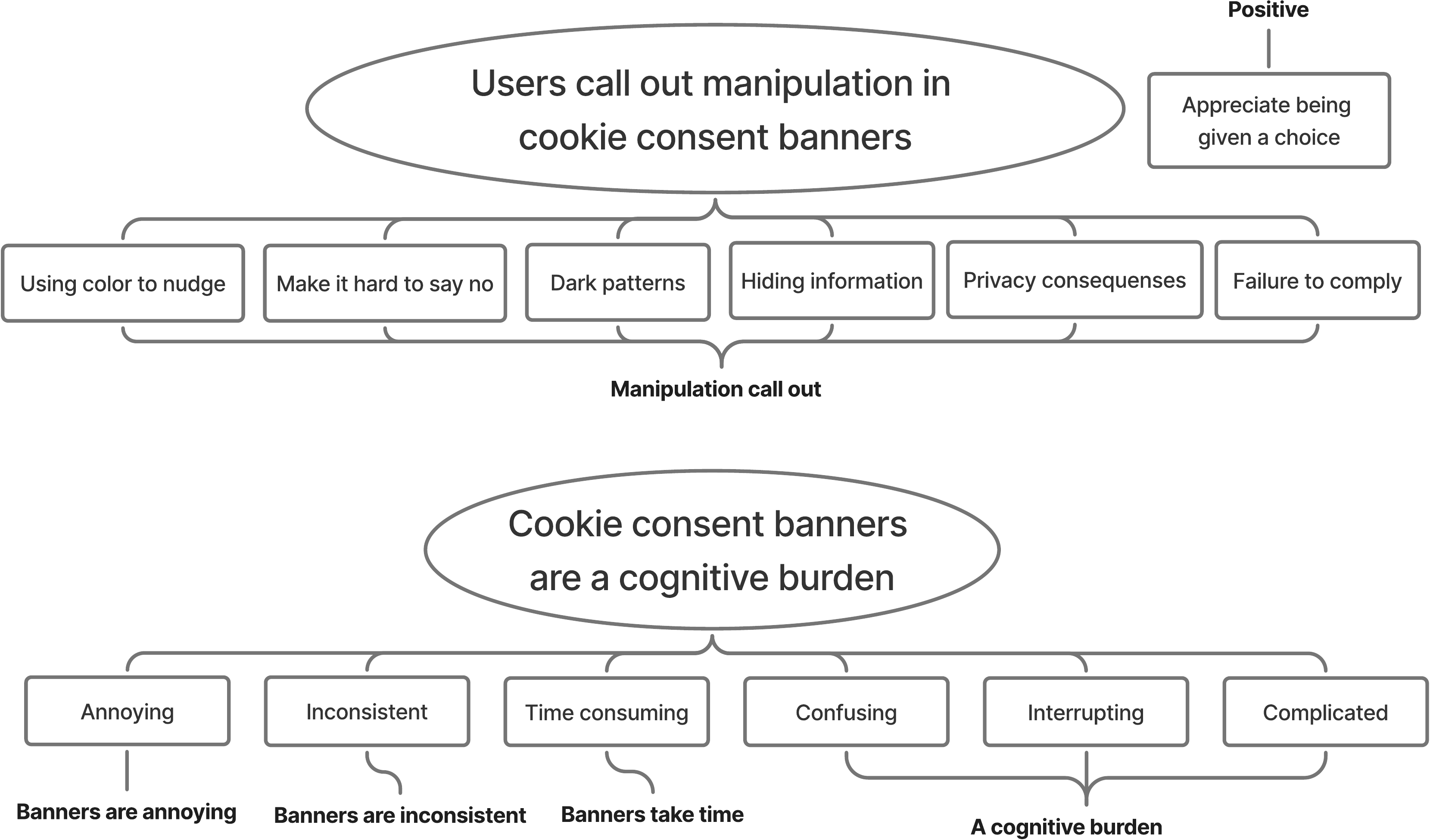}
    \caption[Themes partial consent banners]{\textbf{Revised themes for ``Cookie consent banners are''.} The theme was transformed into three revised themes during phase 4.}
    \label{fig:themes_partial_consent_banners}
\end{figure}